\documentclass[a4paper,onecolumn,12pt]{article}
\usepackage{graphicx}
\usepackage{color}
\usepackage{lscape}
\usepackage{epsfig}
\usepackage{chngpage}
\usepackage{geometry}
\usepackage{psfrag}
\usepackage{pdfrack}
\usepackage{setspace}
\title{An FPGA Based Phased Array Processor for the Sub-Millimeter Array}
\author{Vinayak Nagpal \footnote{Chalmers University of Technology, Gothenburg, Sweden} \\ \textit{advised by} Jonathan Weintroub \footnote{Harvard Smithsonian Center for Astrophysics, Cambridge, MA}}

\begin{document}
\date{September 2005}
\linespread{1.3}
\maketitle
\newpage
\tableofcontents
\newpage
\listoffigures
\newpage
\onehalfspacing
\begin{abstract}
It has been widely acknowledged that Very Long Baseline Interferometry (VLBI) in the submillimeter wavelengths can make imaging observations of super massive black holes possible. The Sub-Millimeter Array (SMA) along with the James Clerk Maxwell Telescope (JCMT) and Caltech Submillimeter Observatory (CSO) on the Mauna Kea summit in Hawaii can together provide a large collecting area as one or more stations for VLBI observations aimed at studying an event horizon. To work as a VLBI station with full collecting area the SMA (or a combination SMA, JCMT, CSO antennas) would need a processor to enable phased array operation. This masters project focusses on building such a processor.
\\

Back end processing for high bandwidth radio telescopes has traditionally been done using custom designed application specific integrated circuits (ASIC). Recent advances in Field Programmable Gate Array (FPGA) technology have made FPGAs both powerful and economically viable for radio astronomy back ends. We have attempted to take advantage of these advances and built a proof-of-concept $500$~MHz phased array processor for the SMA using FPGAs. The phased array processing is done in the time domain using high speed sampling and digital delay lines. The design is capable of spooling the phased sum to a Mark $5b$ VLBI data recorder. It is based on hardware built by the Berkeley Wireless Research Center and the Berkeley Space Science Laboratory.   
\\

We digitize signals after the $1^{st}$ SMA downconvertor using $1024$~MHz sampling and have demonstrated the capability to sum signals from $8$ antennas through programmable digital delay lines up to a precision of $\approx 1/10$ the sampling rate i.e. $0.1$~ns. To calibrate geometric, atmospheric and instrument delays for accurate phasing, a single baseline $512$~MHz $32$ channel FX correlator has also been designed to fit on a single FPGA chip.
\end{abstract}
\newpage
\section{Introduction}
\subsection{Science Goals}
The primary scientific goal of recent work to extend Very Long Baseline Interferometry (VLBI) into the sub-millimeter regime is an imaging observation of the event horizon of a black hole \cite{bh1}. In this context, the sources most likely to be studied are $Sgr A*$ and $M87$.  VLBI at  $0.8$~mm wavelength has the potential to image up to $20$~micro-arc~second angular resolution. There is also a radiative transfer advantage obtained due to reduced electron scattering.  
Therefore we have a strong case to retrofit the Sub Millimeter Array (SMA) with a phased array processor and VLBI recording interface, thereby enabling it to participate in such VLBI observations with its full collecting area.

%Field Programmable Gate Arrays (FPGA) are attractive alternatives to Application Specific Integrated Circuits (ASIC) because of their reduced development time and the option of rapid prototyping. We have employed fast digital sampling and FPGA technology to develop a proof-of-concept phased array processor for the SMA. The design objective was to process $500$~MHz bandwidth of single polarization data from $8$ antennas, compute their phased sum in real time and spool the result to a Mark 5b recording unit. The idea was to architect a workable design out of available state of the art hardware platforms and tools.
%We have used iADC sampling boards based on Atmel ADCs, iBOB boards based on Xilinx FPGAs, all developed by the Berkeley Wireless Research Center (BWRC). BWRC supplies a library of radio astronomy related design blocks to aid development with their boards. We have collaborated with BWRC and extended their library with phased array processing blocks and designed upgrades to other library elements to suit the requirements of our project.
%We digitize SMA data right after the first down conversion stage using a sampling rate of $1024 Msamples/sec$. 

%Our design demonstrates the capability to sum signals from $8$ antennas through programmable digital delay lines up to a precision of $\approx 1/10$ the sampling rate i.e. $0.1~$ns. To calibrate geometric, atmospheric and instrument delays for accurate phasing, a single baseline $512~$MHz $32$ channel FX correlator has also been designed to fit on a single FPGA chip.

\subsection{Project Objectives}
To accommodate the development of a more or less complete system within the purview of a masters thesis it was decided to build a proof-of-concept system which would take IF signals from $8$ antennas. In principle these could include any combination of SMA/JCMT/CSO antennas. We decided to limit ourselves to single polarization data and only $500$~MHz (of the available $2$~GHz) bandwidth. The objective was to compute in real time the phased sum of these antennas and spool the result to a Mark~$5b$ VLBI recording unit and take care of the various calibrations involved. To cut short the design time we decided to use the iBOB FPGA boards and iADC sampling boards built by the Center for Astronomy Signal Processing and Electronics Research (CASPER) \footnote{CASPER works in collaboration with the Berkeley Wireless Research Center, Space Science Laboratory and Radio Astronomy Laboratory at UC Berkeley} group at Univ. of California Berkeley. The time schedule for the project was fixed at $10$ months.

\subsection{Project Partners}
The CASPER \cite{aaron} team at UC Berkeley is working extensively towards developing FPGA based technology for accelerating and standardizing the development of back ends for radio telescopes. The CASPER paradigm focusses on building general purpose FPGA based hardware boards and provide an extensive library of pre designed blocks which can be used to quickly and efficiently design digital subsystems commonly required by radio telescopes. \\
This paradigm was being deployed by the Massachusetts Institute of Technology Haystack Observatory in next generation digital back end for their Mark $5b$ data VLBI storage equipment. The proof-of-concept SMA phased array processor also fitted very well into the capabilities of CASPER boards. In addition it was found that interfacing the processor with VLBI storage equipment would also become very easy if both sub systems used the same hardware platforms. \\ This masters project thus involved extensive collaboration with CASPER in terms of acquiring training/technical support also with MIT/Haystack for interfacing with VLBI storage equipment.

\subsection{The iBOB and iADC Hardware Platform}
\begin{figure}
\begin{center}
\includegraphics[angle=270, width=10cm]{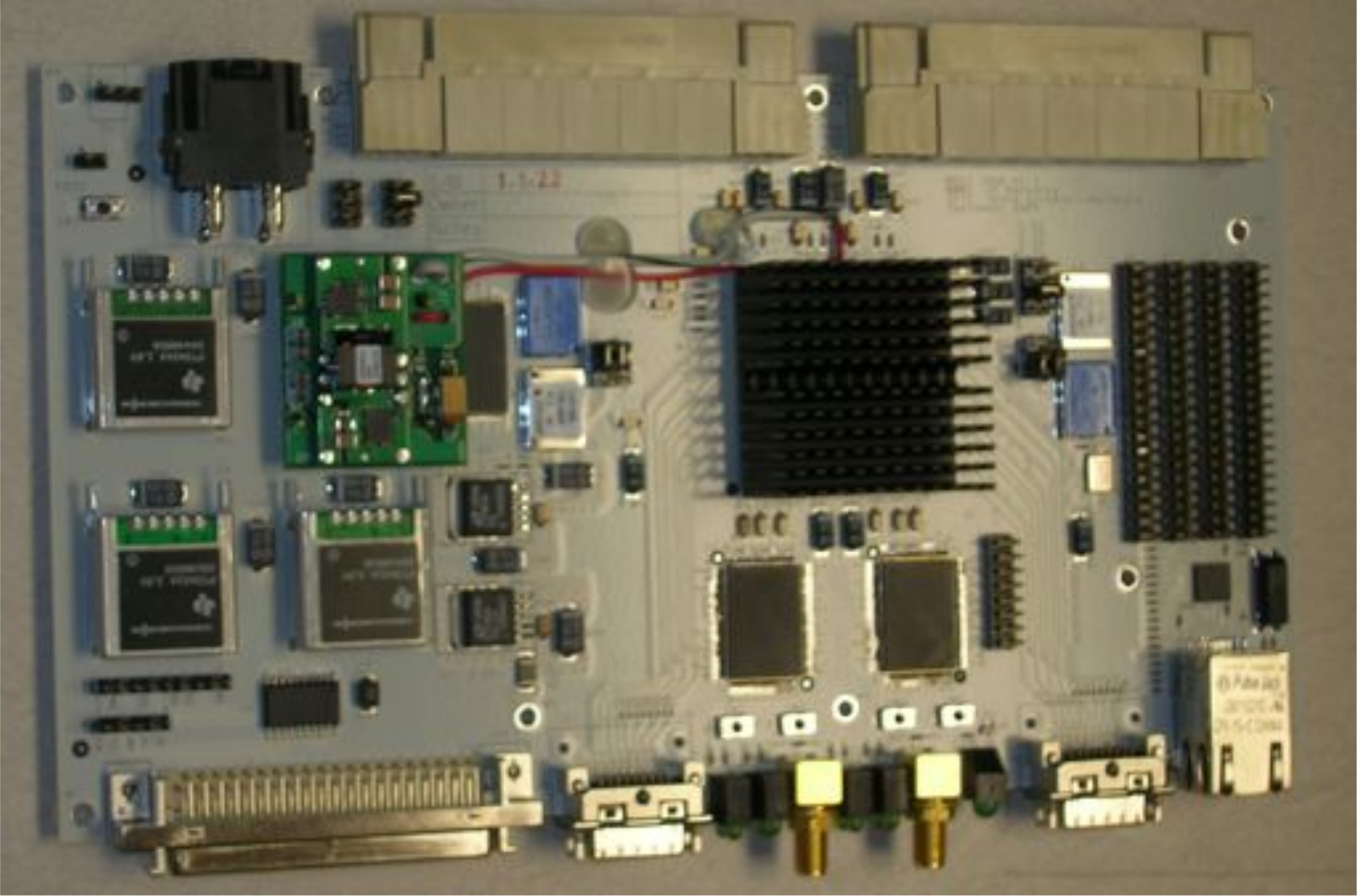}
\end{center}
\caption[ Picture showing iBOB.]{\label{ibobpic}The iBOB board, on the top can be seen 2 ZDOC connectors where iADCs can plug in.}
\end{figure}

\begin{figure}
\begin{center}
\includegraphics[angle=270, width=10cm]{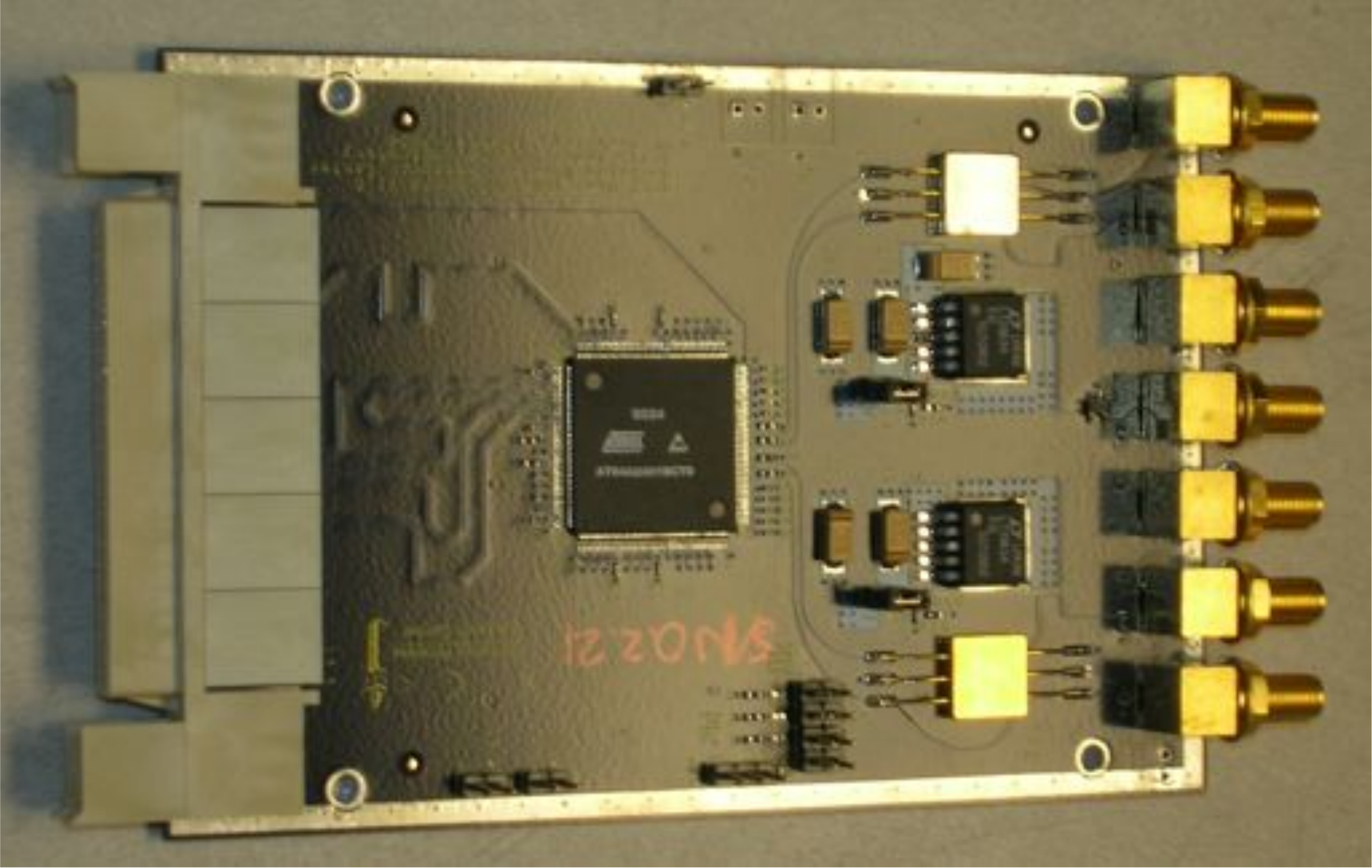}
\end{center}
\caption[Picture showing iADC.]{\label{iadcpic} The iADC board with ZDOC connector on left and analog inputs for two signals, sampling clocks and a synchronization input on right.}
\end{figure}

\begin{figure}
\begin{center}
\includegraphics[width=10cm]{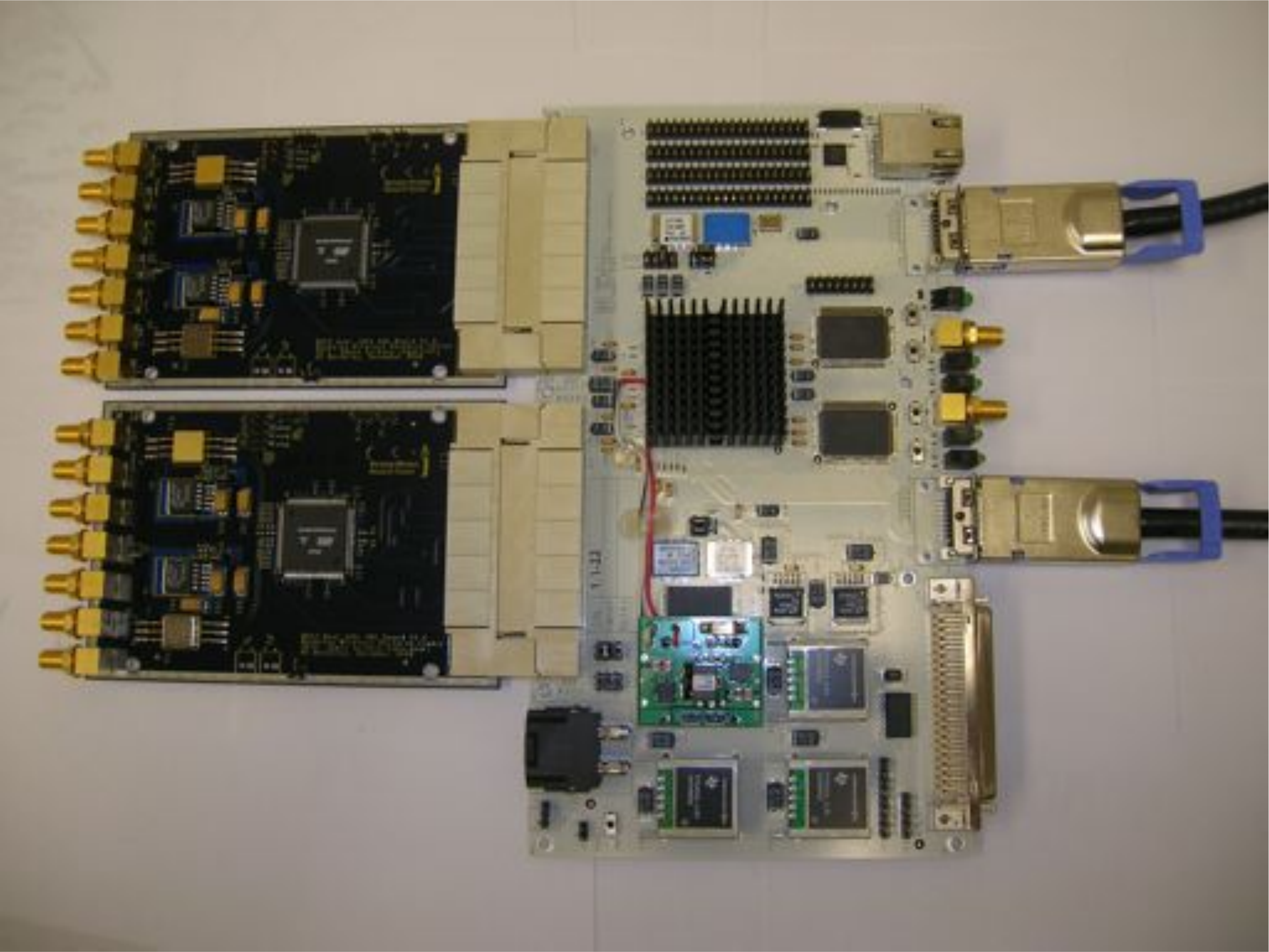}
\end{center}
\caption[Picture showing iBOB with 2 iADCs plugged in]{\label{both} Picture showing iBOB with 2 iADCs plugged in and 2 Infiniband cables connected on the right for streaming data to and from other boards over the XAUI links.}
\end{figure}
\begin{figure}
\begin{center}
\includegraphics[width=4cm]{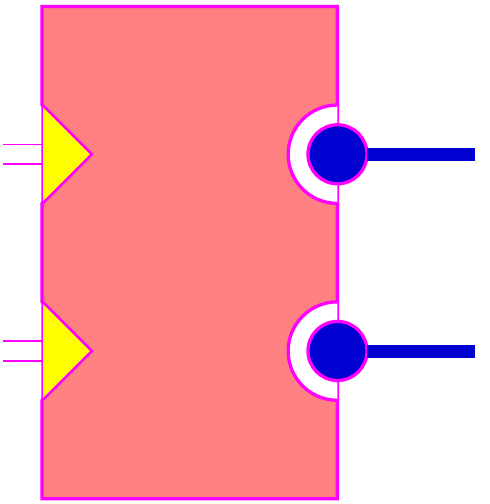}
\end{center}
\caption[Symbolic Diagram for iBOB + 2 iADCs + 2 Infiniband links.]{\label{symbol} Symbolic Diagram for iBOB + 2 iADCs + 2 Infiniband links. The yellow triangles represent iADC boards plugged into the iBOB and blue links represent the infiniband connectors and cables.}
\end{figure}
\begin{figure}
\begin{center}
\includegraphics[scale=0.7]{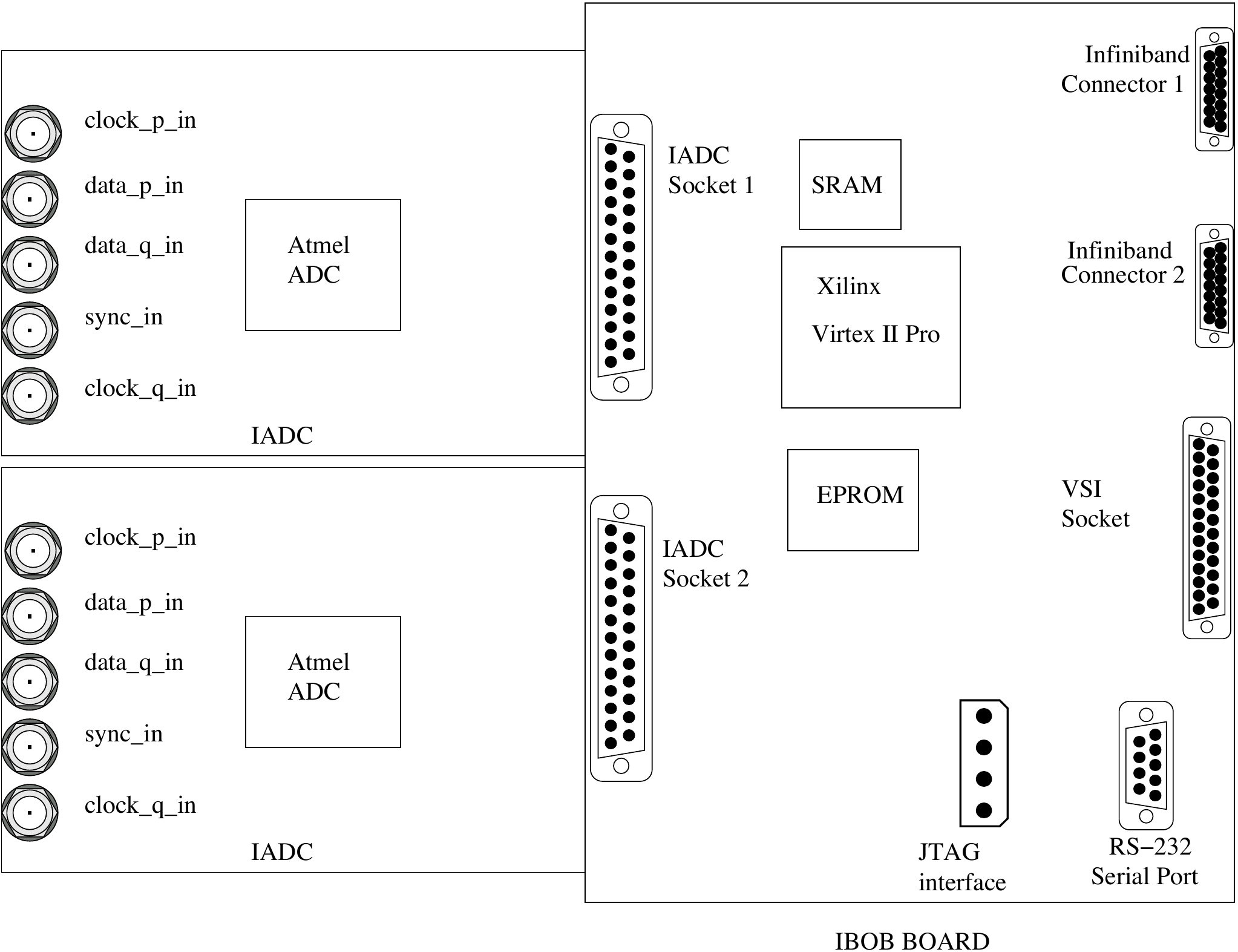}
\end{center}
\caption[Interface Diagram for iBOB + 2 iADCs]{\label{adcdia}Interface Diagram for iBOB + 2 iADCs showing the various analog and digital interfaces on the iBOB setup with iADCs. The connectors shown in this diagram are symbolic only and do not represent the appearance or pin-out of actual connectors.}
\end{figure}

The iBOB and iADC boards are a part the BEE2 (Berkeley Emulation Engine) FPGA platform. The BEE2 platform was developed at the Berkeley Wireless Research Center(BWRC) primarily for applications requiring multi tera-flops of processing power and for emulating multi-processor computer architectures. 
For the purpose of this project we have used the iBOB and iADC boards which are add-ons to the BEE2 suite.
The iBOB is equipped with a Xilinx Virtex II Pro (vp50) FPGA and high speed data interfaces (Infiniband connectors). The iADCs are smaller boards which plug directly onto the iBOB and provide fast sampling using an Atmel analog to digital conversion chip. A single iADC can provide $2$~GHz sampling for one channel or $1$~GHz sampling for 2 data channels. One iBOB can mount 2 such iADC boards. Figure \ref{ibobpic} shows a photograph of one iBOB, Figure \ref{iadcpic} shows an iADC and Figure \ref{both} shows a iBOB $+ 2$ iADC setup. The various interfaces of this setup are shown in Fig. \ref{adcdia}.
Brief descriptions of various components and interfaces shown in the diagram are listed below and Figure \ref{symbol} shows symbolic representation of a iBOB$+ 2$ iADC$+ 2$ Infiniband setup.

\begin{enumerate}
\item RS-232 Serial Port: This is used to communicate control instructions to FPGA design from an external computer. 
\item JTAG Connector: (Joint Test Action Group) JTAG interface is used for loading designs into the FPGA or for burning the EPROM which stores a default design loaded into the FPGA at power-on. In this project we have cascaded 3 iBOBs into a single JTAG chain to access all FPGAs using one programming cable.
\item Infiniband Connectors: Each iBOB is equipped with 2 Infiniband connectors. Infiniband is a high speed bi-directional serial bus. The BEE2 platform uses these in a $10$~Gbps data rate configuration (Infiniband supports maximum of $120$~Gbps). The interface derives its clock from a $156$~MHz crystal oscillator provided on-board. \emph{Rocket IO} components available within the Virtex II Pro provide the physical layer to drive these links. The transport protocol deployed is based on the IEEE 802.3ae $10$~Gb Ethernet specification also called \textit{X} (Roman Numeral $10$) \textit{Attachment Unit Interface} or \emph{XAUI}. The BEE2 platform uses a proprietary XAUI core licensed from Xilinx. 
\item VSI Connector: The Versatile Scientific Interface (VSI) bus is the standard interface adopted for VLBI and directly plugs into the Mark $5b$ data storage modules. The VSI interface logic is designed into the FPGA as a BWRC library component.
\item iADC: The Atmel (AT84AD001) sampling chip can be configured from the FPGA. The \texttt{clock-in} ports on the iADC are driven with a $-6$~dBm sine wave clock of $1024$~MHz. This derives the FPGA clock using a divide-by-4. The FPGA processes 4 data samples per clock (demux-by-4) at a clock rate of $1024/4=256$~MHz. The \texttt{sync-in} inputs allow a synchronization pulse to provide alignment markers for multiple iBOB designs. The circuitry to interface with and operate the ADC from the FPGA is provided by BWRC as a standard library component.
\item SRAM: The FPGA can access a SRAM chip in addition to its own memory for off-chip storage. 
\end{enumerate}

\subsection{Development Platform} 
\begin{center}
\begin{figure}
\includegraphics[scale=0.6]{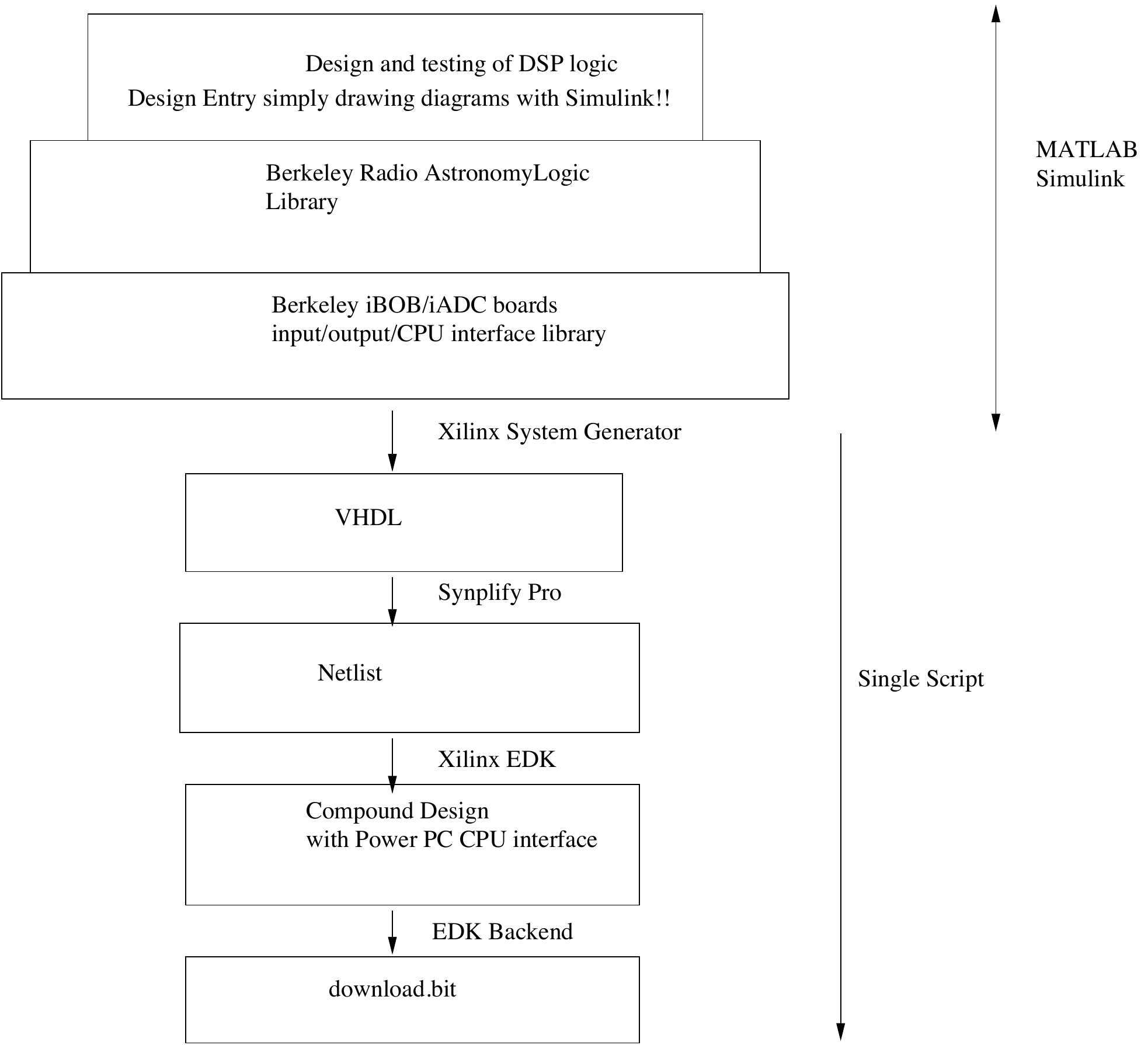}
\caption[Summary of BEE2 development platform]{\label{toolflow} The BEE2 development workflow is based in MATLAB Simulink which is used for design entry simulation and verification. To assist the design CASPER provides a library of highly optimized and parametrize-able blocks for signal processing required frequently in radio astronomy applications. Compilation of designs to VHDL, netlist, integration of embedded software etc are all done through a single script built into the CASPER workflow.}
\end{figure}
\end{center}

Complementing the BEE2 hardware platform is an improved FPGA development environment. The board specific details are masked from the logic designer by providing highly parameterized library components for all input/output interfaces. The BEE2 development platform is summarized in Figure \ref{toolflow}.
\begin{figure}
\begin{center}
\includegraphics[width=10cm]{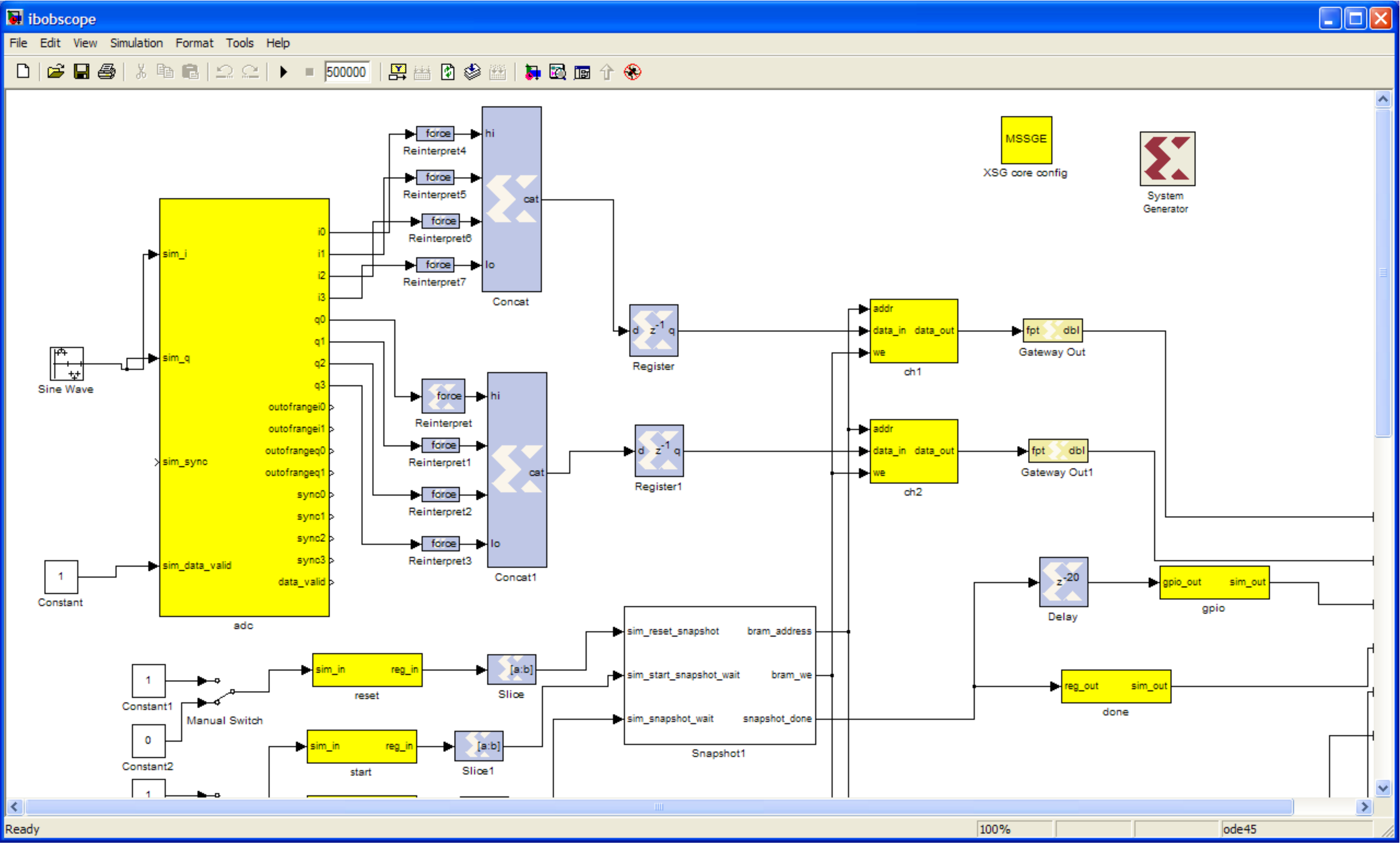}
\end{center}
\caption[Simulink Screenshot.]{\label{simulink} A screenshot of Simulink based design. The yellow blocks are models of external interfaces like the iADC or software accesible registers, the blue blocks are Xilinx supplied hardware primitives.}
\end{figure}

\subsubsection{DSP Logic Design} The DSP logic design is done using the Xilinx System Generator Block set for MATLAB/Simulink. This reduces the logic design task to that of drawing diagrams using basic hardware building blocks which can be simulated and tested in the Simulink environment. The designs can be later compiled into synthesize able VHDL using Xilinx System Generator. A screenshot of Simulink screen with Xilinx components (blue blocks) is shown in Figure \ref{simulink}.
\subsubsection{Interface Logic Design} The FPGA interfaces like iADC, SRAM, PPC (refer item \ref{PPC} below) registers, PPC Shared RAM, XAUI connectors, VSI bus and general purpose I/O (GPIO) units like LEDs, switches are done using BWRC library components which are in the form of Simulink blocks. These blocks serve a dual purpose. They provide a Simulink model of the interface for the logic design and testing phase and also provide a logic circuit to take care of the actual interfacing details during the compilation/synthesis phase. The logic circuit replaces the Simulink model automatically if the design compilation is invoked using BWRC scripts. Logic interface blocks can be seen as yellow blocks in Figure \ref{simulink}.
\subsubsection{Embedded Software} \label{PPC} Virtex II Pro offers two on chip 32-bit Power PC (PPC) processors in addition to the reconfigure-able fabric. The FPGA can be configured to bridge the memory and data busses of these processors to the clock domain of the designer's custom DSP logic. This enables the PPC to interact with DSP logic by read/write operations on certain registers/RAMs. BWRC has designed a small operating system to run on these PPCs called \emph{Tiny Shell}. Tiny Shell provides a basic command interpreter and  serial port driver which is used to communicate with an external computer for sending control information.
\subsubsection{Design Synthesis} Compilation scripts (provided by BWRC) take care of the underlying details of replacing Simulink models with interface circuitry. They also generate the required constraint files, embedded software configurations and invoke various back-end tools automatically with the required parameters. In effect the compilation of Simulink designs into VHDL, then to net-lists, mapping of busses in Xilinx Embedded Design Kit (EDK) and back-end synthesis are all reduced to a single click operation.

\subsection{The Submillimeter Array and IF/LO subsystem}
The Sub-MM Array (SMA) \cite{sma} is an $8$ element interferometer operating in the range of $200-900$~GHz on the Mauna Kea summit in Hawaii. Each element is a steerable smooth parabolic reflector antenna having a diameter of $6$~m. The antennas can be moved between pads to provide different size array configurations. The longest baselines obtainable are about $0.5$~km. The total bandwidth available with SMA receivers is $2~$GHz. Our phased array processor is proposed to tap into the SMA signal chain after the first down converter which presents $1~$GHz signal bandwidth centered at about $1~$GHz. We sample this data at a rate of $1024~$MHz allowing a Nyquist bandwidth of $512~$MHz. 
%Before digitizing we use $480~$MHz wide anti-aliasing chunk filters centered at $760~$MHz or $1280~$MHz respectively to choose either of two available $512~$MHz bands. The choice of $1024~$MHz for sampling was driven by the fact that Mark 5b data recording equipment requires its data rate (in MHz) to be $1024$~Msamples/sec. The choice two available $500$MHz IF bands are shown in Figure \ref{iflo}.

\section{The SMA Phased Array Processor}
\subsection{Generic Phased Array}
\begin{figure}
\psfrag{A}{$\tau (\lambda_{sky})$}
\psfrag{K}{$\tau (\lambda_{if})$}
\psfrag{B}{$\phi$}
\psfrag{f}{$f$}
\psfrag{C}{Mix}
\psfrag{D}{LO}
\psfrag{E}{LO}
\psfrag{F}{$\phi_1$}
\psfrag{G}{$\phi_2$}
\psfrag{H}{$\phi_1 - \phi_2$}
\begin{center}
\pdfrackincludegraphics[scale=0.5]{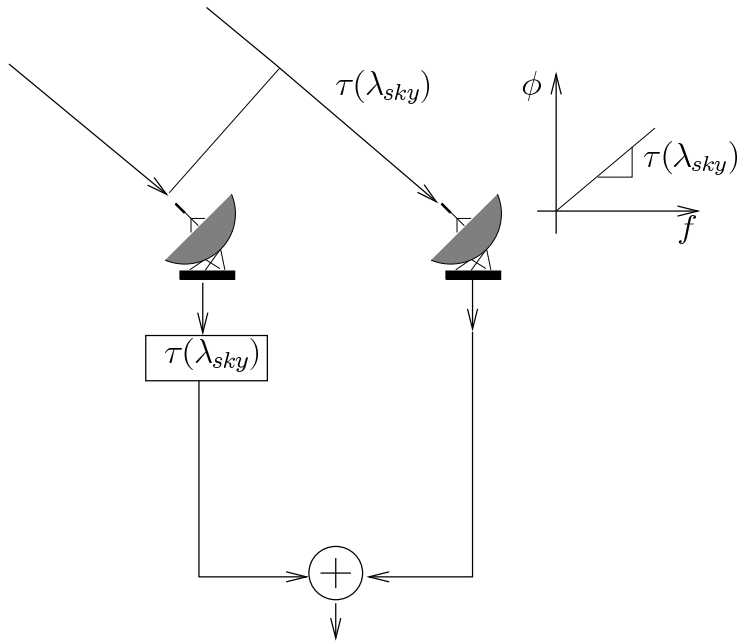}
\end{center}
\caption{\label{s2pa} A simple two element phased array without heterodyning. The geometric delay $\tau$ is caused and corrected at the sky frequency $\lambda_{sky}$}
\end{figure}

\begin{figure}
\psfrag{A}{$\tau (\lambda_{sky})$}
\psfrag{K}{$\tau (\lambda_{if})$}
\psfrag{B}{$\phi$}
\psfrag{f}{$f$}
\psfrag{C}{Mix}
\psfrag{D}{LO}
\psfrag{E}{LO}
\psfrag{F}{$\phi_1$}
\psfrag{G}{$\phi_2$}
\psfrag{H}{$\phi_1 - \phi_2$}
\begin{center}
\pdfrackincludegraphics[scale=0.5]{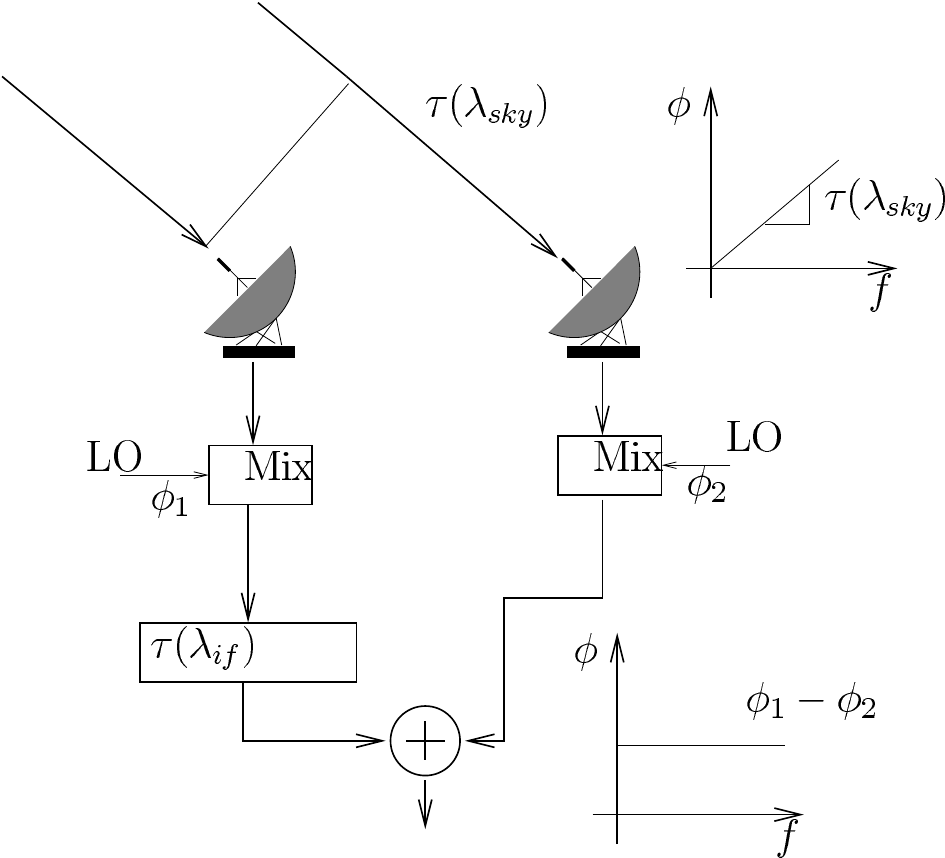}
\end{center}
\caption{\label{h2pa} A more realistic heterodyne two element phased array where the geometric delay $\tau$ is caused at the sky frequency $\lambda_{sky}$ and compensated after down-conversion at the intermediate frequency $\lambda_{if}$. The offset caused due to LO phase mismatch also remains. Both these effects need to be corrected for accurate phasing.}
\end{figure}

A phased array is a group of antennas in which the relative phases of the respective signals received at the antennas is varied in such a way that the voltage sum of these signals causes the effective radiation pattern of the array to be reinforced in a desired direction. For a single baseline interferometer a simple phased array can be seen in Figure \ref{s2pa}. The phase plot shows the phase difference between the two received signals as a function of frequency. 
A delay $\tau$ corresponds to the slope $\frac{\phi}{f}$ in the phase plot. In a heterodyne receiver the sky signal is down converted to an intermediate frequency as shown in Figure \ref{h2pa}. This introduces an effect if the local oscillator phases at two antennas is slightly different. As seen in the phase plot this causes a flat phase shift over the entire frequency band corresponding to $\phi_1 - \phi_2$. This effect cannot be corrected using delay elements alone. If this effect is not corrected in the analog domain by carefully adjusting LO phases the backend needs to use frequency domain techniques to correct for this phase shift. The SMA first LO's are phase programmable and the correlator software can adjust the LO phases such that $\phi_1-\phi_2=0$. 
The $\tau$ delay compensation must be done after down conversion and is not exact because the compensation is done at a higher wavelength $\lambda_{if}$ 
instead of the actual $\lambda_{sky}$. This can be corrected during fringe rotation explained in section \ref{fringerot}.

\begin{figure}
\psfrag{A}{$\tau (\lambda_{sky})$}
\psfrag{K}{$\tau (\lambda_{if})$}
\psfrag{B}{$\phi$}
\psfrag{f}{$f$}
\psfrag{C}{Mix}
\psfrag{D}{LO}
\psfrag{E}{LO}
\psfrag{F}{$\phi_1$}
\psfrag{G}{$\phi_2$}
\psfrag{H}{$\phi_1 - \phi_2$}
\begin{center}
\pdfrackincludegraphics[scale=0.5]{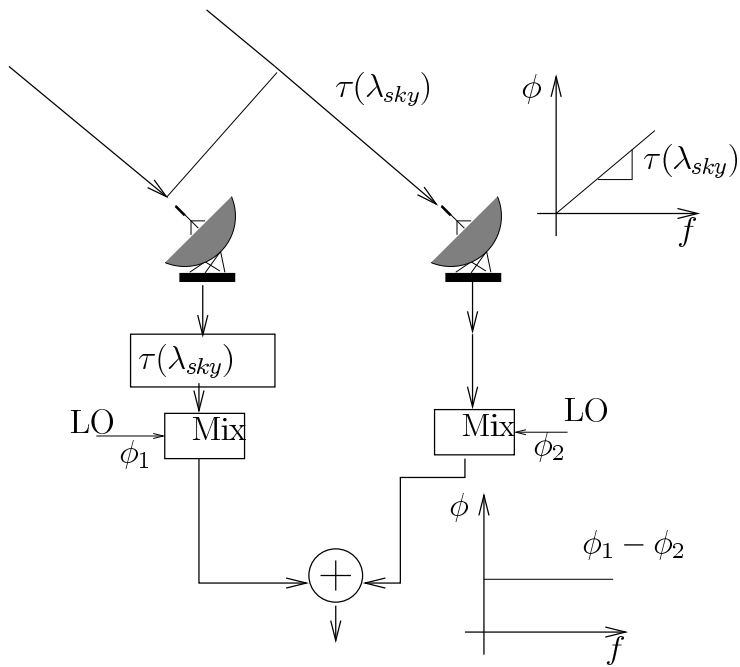}
\end{center}
\caption{\label{sh2pa} A simple two element heterodyne phased array (assuming we have the capability to compensate for delays before down conversion). During down-conversion the difference in LO phases causes a constant phase offset across the entire band which needs to be corrected.}
\end{figure}
The delay line compensation can be achieved using two equivalent approaches.
\subsubsection{Time Domain Phased Array Processing}

\begin{figure}
\psfrag{A}{Delay $\tau_1$}
\psfrag{B}{Delay $\tau_2$}
\psfrag{C}{Delay $\tau_3$}
\psfrag{D}{Delay $\tau_n$}
\begin{center}
\pdfrackincludegraphics[width=9cm]{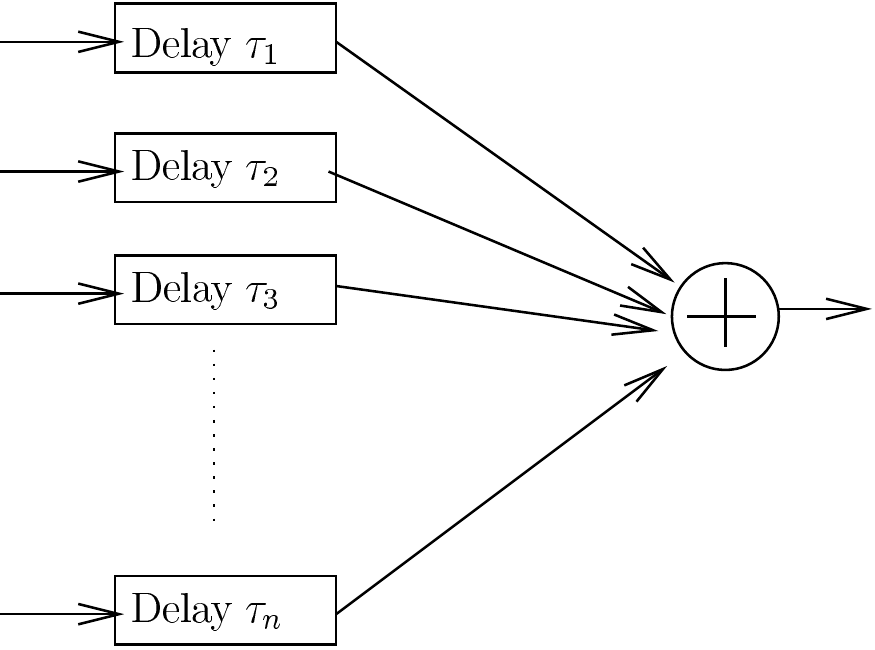}
\end{center}
\caption{\label{td} A simple time domain beamformer with $n$ channels. A channel $i$ is delayed in time by $\tau_{i}$ such that the sum is phased to give highest gain in a particular direction.}
\end{figure}
Figure \ref{td} shows the time domain approach which utilizes a variable delay line per antenna. This simple approach can compensate delays only (not phase offset effects). The accuracy or phase coherence performance of the time domain beamformer depends on the delay step size. The smaller the delay step $\tau_{min}$ is compared to the signal bandwidth $B$ the better will be the phase coherence of the beam formed signal.
\begin{equation}
\tau_{min} << \frac{1}{2B}
\end{equation}
\subsubsection{Frequency Domain Phased Array Processing}

\begin{figure}
\psfrag{A}{FFT $N$}
\psfrag{B}{$\phi$ adjust}
\begin{center}
\pdfrackincludegraphics[width=9cm]{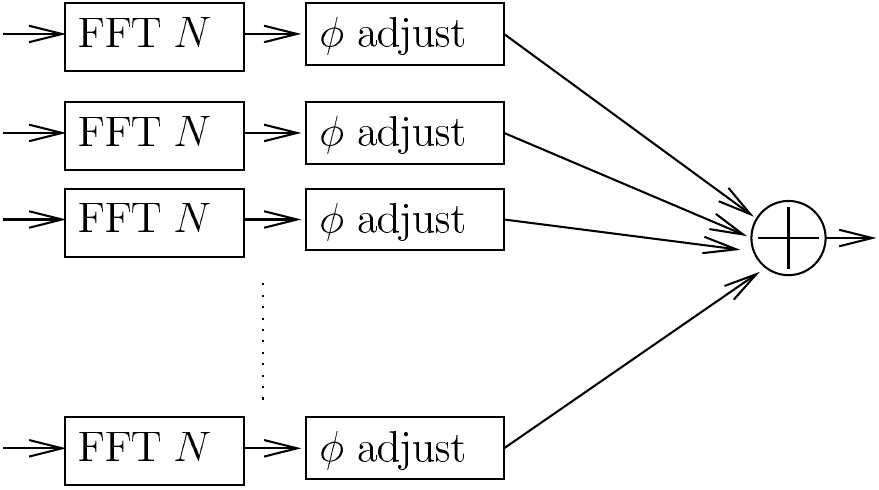}
\end{center}
\caption{\label{fd} A Frequency Domain Beamformer where the $n$ channels are delayed in frequency domain by choosing a correct phase offset $\phi$ for each frequency component such that the combined spectrum for the channel shows a slope corresponding to desired delay $\tau_{i}$. The sum must be converted back to time domain by an inverse Fourier transform to be equivalent to the time domain beam former. In our application since the signal is stored in frequency domain in the Mark 5B storage unit we can omit this inverse transformation.}
\end{figure}
Figure \ref{fd} shows frequency domain delay lines. The $\phi$ adjust blocks adjust the channel phase such that the phase response will have the appropriate delay slope. In addition to delays this approach can compensate for fixed and variable phase offsets in the system like the LO phase difference described earlier. The phase coherence of the frequency domain beamformer improves by increasing the number of frequency channels $N$. A big disadvantage of this approach is the need to have multiple (1 per antenna) FFT and Inverse FFT blocks which are computationally expensive. For the SMA phased array processor we have chosen to use the time domain approach because the phase offset effects can be corrected using existing SMA analog subsystems.

\subsection{System Architecture}
\begin{figure}
\psfrag{A}{M5 Recorder}
\psfrag{B}{DBE}
\psfrag{T}{VSI}
\psfrag{C}{VLBI In}
\psfrag{U}{iBOB-1}
\psfrag{V}{iBOB-2}
\psfrag{P}{Ant 1}
\psfrag{Q}{Ant 2}
\psfrag{R}{Ant 3}
\psfrag{S}{Ant 4}
\psfrag{E}{Ant 5}
\psfrag{F}{Ant 6}
\psfrag{G}{Ant 7}
\psfrag{H}{Ant 8}
\psfrag{X}{$8$~Gbps}
\psfrag{Y}{$8$~Gbps}
\psfrag{Z}{Infiniband}
\pdfrackincludegraphics[width=10cm]{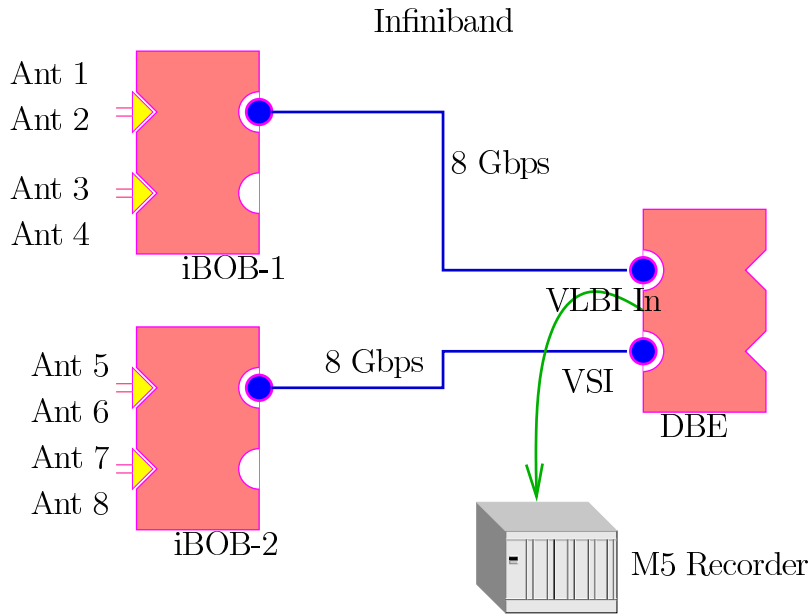}
\caption{\label{symarch}Symbolic System Architecture: The blue lines correspond to infiniband links, the yellow triangles correspond to iADC boards.}
\end{figure}

The SMA phased array processor is built upon building blocks described in the previous sections. 
A system architecture diagram showing various components and interconnects is shown in Figure \ref{symarch}.
%\begin{figure} 
%\includegraphics[scale=0.6]{arch}
%\caption{\label{arch} Overall System Architecture}
%\end{figure}
The entire system can be implemented on $4$ iBOBs and $4$ iADC boards. The first $2$ iBOBs implement identical designs that implement digital delay lines and addition logic. The $3^{rd}$ iBOB implements the digital back-end \emph{DBE} for Mark 5b storage unit and the 4th iBOB is used for system calibration. The symbolic representation for the system (without calibration) is seen in Figure \ref{symarch}. 

\subsection{Analog Subsystem}
\begin{figure}
\psfrag{A}{$0.5$~GHz}
\psfrag{B}{$1.5$~GHz}
\psfrag{C}{$1$~GHz}
\psfrag{D}{SMA $1^{st}$ DCV}
\psfrag{E}{Block Filters (MHz)}
\psfrag{F}{$528$}
\psfrag{G}{$768$}
\psfrag{H}{$1008$}
\psfrag{I}{$1040$}
\psfrag{J}{$1280$}
\psfrag{K}{$1520$}
\psfrag{L}{$1024$~MHz}
\psfrag{M}{$M$}
\psfrag{f}{$f$}
\begin{center}
\pdfrackincludegraphics[scale=0.6]{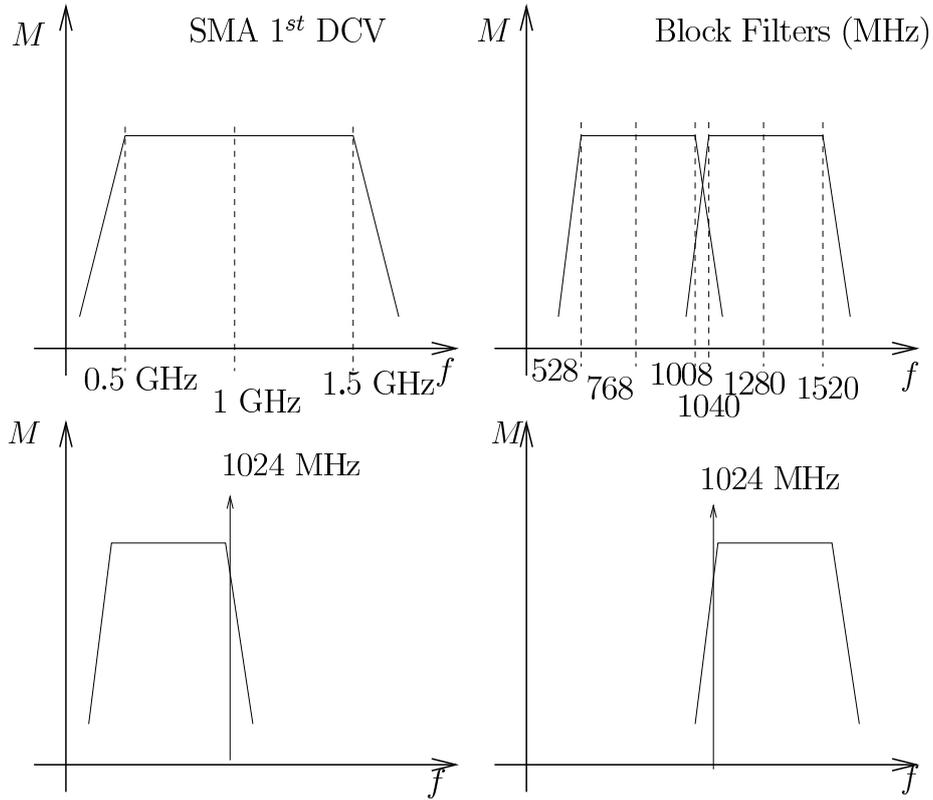}
\end{center}
\caption{\label{iflo} SMA IF/LO Subsystem: Top left figure shows the output of the $1^{st}$ down-converter. Top right shows the bandpass of 2 block filters we use for antialiasing. The two bottom figures show the filtered bands which we feed to the iADC for sampling. We can choose a band of interest by deploying the appropriate filter.}
\end{figure}
\subsubsection{IF/LO Subsystem} Figure \ref{iflo} shows the distribution of bandwidth at the IF input to ADCs. The SMA $1^{st}$ downconverter output presents $1$~GHz of bandwidth centered at $\approx 1$~GHz. We can choose either of two $500$~MHz bands from this using any one of two block filters which were custom ordered for this purpose. The available Nyquist bandwidth with $1024$~MHz sampling is $512$~MHz, we utilize most of this by deploying anti-aliasing filters having half-power bandwidth of $480$~MHz centered at $760$~MHz or $1280$~MHz at the IF output. The filter choice determines which band we select as shown in Figure \ref{iflo}. It can be seen that the sampling clock appears on opposite edges of the two available bands causing the band to show a frequency flip in one configuration. The choice of $1024$~MHz for sampling was dictated by the fact that the MIT/Haystack Mark $5b$ recording subsystem is designed to work at this rate.
\subsubsection{Phase Switching} Periodic switching of LO phase is used routinely in most interferometers to ensure that the interferometer output is zero when the input signals have no correlation. The SMA has two levels of phase switching and atleast the $180^\circ$ switching is essential for proper operation of the SMA correlator \cite{colin}. 
Submillimeter receivers cannot have low noise amplifiers and side band separating filters working at the sky frequency in their front end, their first stage is usually a mixer. In a regular heterodyne receiver front end LNA's and filters (at RF) allow easy separation of the lower and upper sidebands, however as submillimeter receivers have a mixer in their first stage alternate methods are used for sideband separation. In the SMA this is achieved using $90^\circ$ phase switching and appropriate addition/subtraction to cancel out one of the sidebands. In addition to this $180^\circ$ phase (Dicke) switching is used for cancelling offset and leakage effects. These two switching functions are achieved using a fast $90^\circ$ Walsh sequence based phase switch superimposed by an independent slow $180^\circ$ Walsh sequence based phase switch. Their combined effect causes the LO phase to switch between $90^\circ$, $180^\circ$, $270^\circ$ and $360^\circ$ based on two superimposing Walsh sequences. In the SMA correlator these phase switches are corrected for and sidebands are separated post correlation. However for the purpose of phased array operation, $90^\circ$ phase switching correction in time domain prior to correlation would require digital real time implementation of $500$~MHz Hilbert Transformers. These are complex to design and computationally expensive. If $90^\circ$ phase switching is turned off, both sidebands would appear overlapped in the signal. For VLBI purposes the unwanted sideband will act like noise and eventually be eliminated in the VLBI correlation processing. This causes a significant Signal-to-Noise Ratio (SNR) disadvantage. In VLBI experiments many single dish telescopes are expected to participate and single dish submillimeter telescopes usually do not have a mechanism to separate the sidebands. Thus in submillimeter VLBI the SNR disadvantage of overlapped sidebands is more or less unavoidable. In this project we plan to run the SMA with $90^\circ$ phase switching turned off for phased array VLBI operation. The $180^\circ$ phase switching can be left running and compensated for easily in the phased array processor by switching between addition and subtraction on respective Walsh ticks.
\subsubsection{Fringe Rate Correction} \label{fringerot} Over time, when observing a point source the output of a interferometer should remain constant. However as the earth rotates, the source appears to move across the sky. This causes the phase difference between the signal received by two antennas to change slowly. This change can be analyzed as the relative doppler shift caused at the two antennas by earths rotation. This effect needs to be compensated for by introducing appropriate phase shifts per antenna before correlation or beamforming. This is called fringe rotation \cite{phd} and can be achieved either digitally by programming correct phase shifts or in analog by shifting the local oscillator (LO) phase with the tracking rate given by equation \ref{fringerate}.
\begin{equation}
\label{fringerate}
\frac{d\Delta \phi}{dt}=\frac{2\pi d \cos(\theta)}{\lambda_{sky}}\frac{d\theta}{dt}
\end{equation}
where $\Delta \phi$ is the relative phase difference to be introduced in a pair of antennas, $\theta=90-e$ where $e$ is the source elevation. The fringe rate can be modified to also correct for the fact that delay compensation is done at the IF rate instead of the RF. (Using a tracking rate adjustment for $\lambda_{if}$). \cite{phd}
In the SMA these effect are corrected for in the RF by the correlator software with phase and frequency agile LOs. (We use the first LO and since we sample before the second LO stage we need to simulate the second LO by the sampling clock). It is therefore not needed to include special provision in the phased array processor to correct for fringe rotation.

\subsection{Mark $5b$ Digital Backend (DBE)} 
\begin{figure}
\begin{center}
\includegraphics[width=8cm]{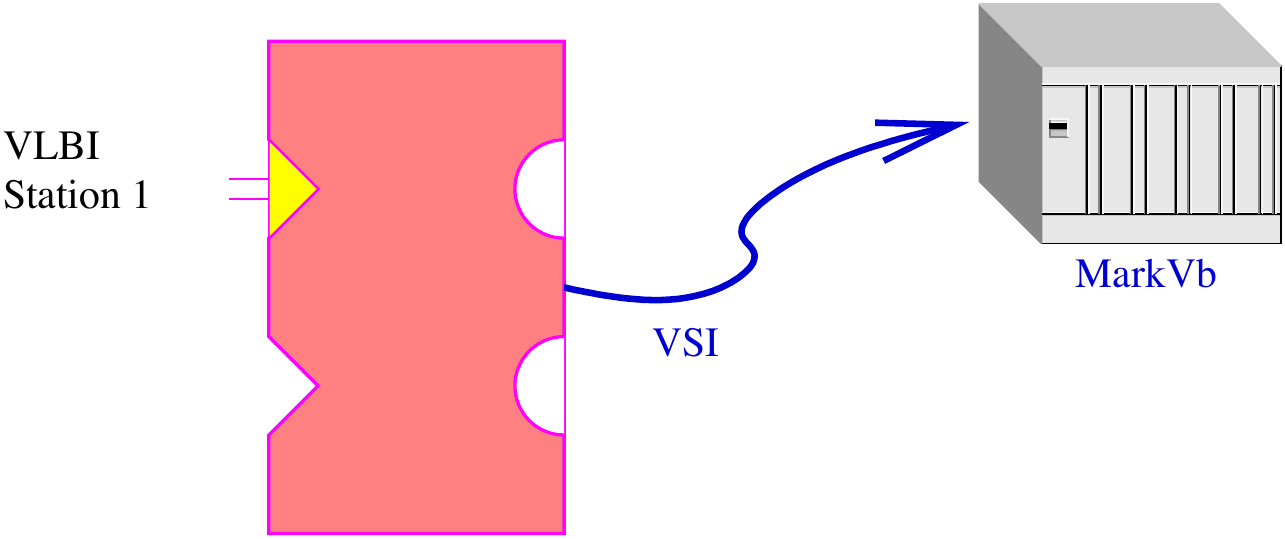}
\end{center}
\caption{\label{dbesym} Mark $5b$ DBE Symbolic System Diagram}
\end{figure}

\begin{figure}
\begin{center}
\includegraphics[width=11cm]{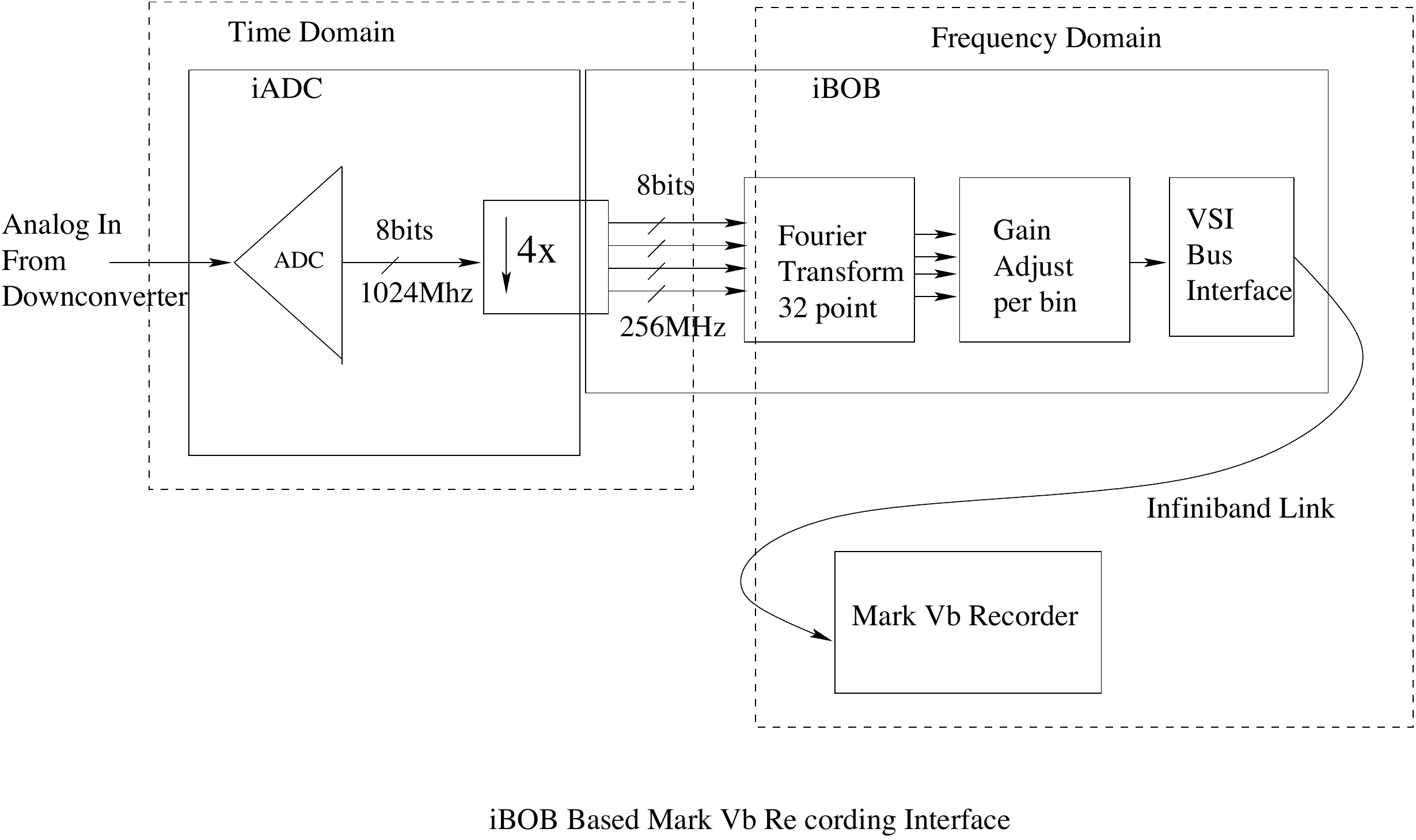}
\end{center}
\caption[Mark $5b$ DBE Block Diagram]{\label{dbe} Mark $5b$ DBE Block Diagram: iBOB and iADC are used to sample $1$~GHz of bandwidth, frequency channelize it and send it over VSI after some conditioning. (This subsystem is built by the CASPER group at UC Berkeley)}
\end{figure}

The DBE is designed by the CASPER group using iBOB boards for MIT/Haystack. The DBE is designed to accept a single channel of $512$~MHz bandwidth at the iADC input. It channels this signal into $32$ frequency bins using poly phase filter banks and fast fourier transforms, it then takes care of compensating for filter bandshapes performs bin weighting and spools the data over the Versatile Scientific Interface (VSI) bus to a Mark $V$ data recorder. Figures \ref{dbesym} and \ref{dbe} show the DBE design system diagram and blocks diagram respectively.
We modified this DBE design (Simulink based) to receive data over XAUI links and sum together the partial phased sums from $2$ iBOBS as shown earlier in Figures \ref{arch} and \ref{symarch}. This results in a $8$ channel phased sum passing through the signal processing chain required for Mark $5b$ data storage.

\subsection{Digital Delay Lines} The digital delay lines are implemented across $2$ iBOB boards. They read sampled data from $8$ antennas using $2$ iADC boards per iBOB, i.e. they receive data at the rate of $1024~$MHz $\times$ $8~$bits$=8~$Gbps $\times$ $4~$channels $\times 2$ boards $=64~$Gbps. Each iBOB implements $4$ digital delay lines each of which is capable of introducing accurate programmable time delays. The maximum supported delay is $4000~$ns and the precision step is $0.1~$ns, which is $1/10$ times the sampling period $T_s$.
\begin{equation}
T_s=\frac{1}{1024MHz}=0.976\times10^{-9}\approx1ns
\end{equation}
\begin{equation}
\tau_{min}=\frac{1}{10}\times\frac{1}{2B}=\frac{1}{10}\times\frac{1}{2\times512\times10^6}=0.0976\times 10^{-9}\approx0.1ns
\end{equation}
\begin{figure}
\begin{adjustwidth}{-4em}{-4em}
\begin{center}
\includegraphics[scale=0.45]{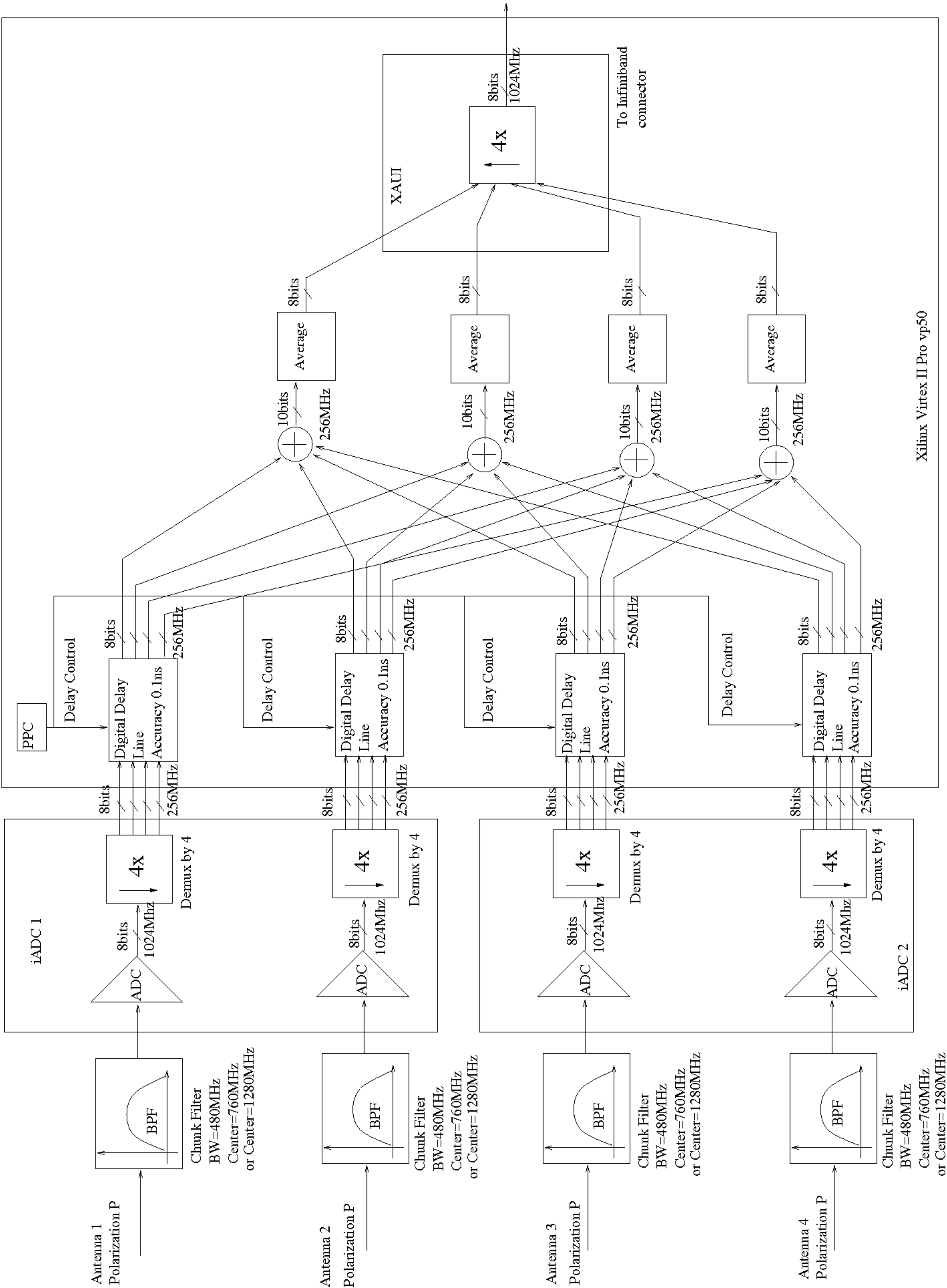}
\end{center}
\caption{\label{beamform} Architecture of 1 Delay Line iBOB Design}
\end{adjustwidth}
\end{figure}
Figure \ref{beamform} shows a detailed design of one of the iBOBs. Since the FPGA cannot be clocked at rates as high as $1024$~MHz the iADC and iBOBs perform a demux-by-4 which presents $4$ samples per clock at a rate of $256$~Mhz.
The design expects to be provided with delay values from the control computer. If delays are correctly programmed the 4 channels are phase aligned. This result is transmitted over XAUI links to the DBE at $8~$Gbps, i.e. $1024$Msamples/sec and $8$bits/sample. All components of this design have been developed under the purview of this masters project. The delay precision of $0.1$~ns is achieved in 3 steps.
\subsubsection{Coarse Delay}
\begin{figure}
\begin{center}
\includegraphics[width=10cm]{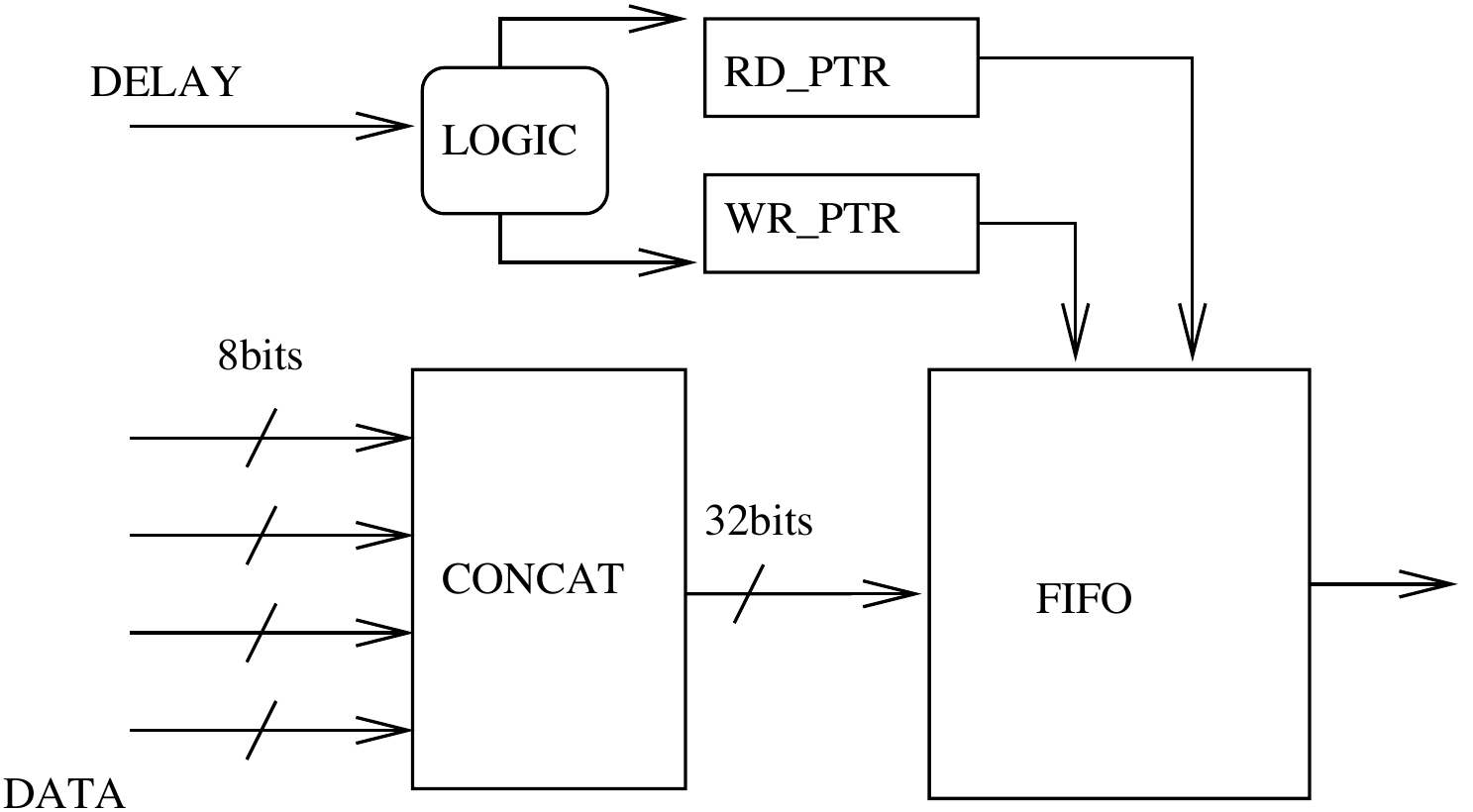}
\end{center}
\caption[Coarse Delay]{\label{coarse} Coarse Delay is implemented using a FIFO like system.}
\end{figure}
The coarse delay achieves a delay step size of $4$~ns and a maximum delay of $4000$~ns. The demux-by-4 is disregarded in this stage. The $4$ $8$-bit samples appearing per clock are concatenated into one $32$ bit number per clock. These are fed into a RAM based FIFO (First In First Out) structure with read and write pointers maintained in flip flops. The FIFO can accommodate $1000$ $32$-bit values i.e. $4000$ time samples. We have designed control logic that controls the read and write pointers based upon the programmed coarse delay value $C$. The control logic is unable to handle coarse delay values smaller than $3$, thus that is used as a base value in all channels. When the programed value is reduced by $1$ the control logic skips a read for one clock by outputting the same word twice. When the programmed value is increased by $1$, the control logic causes a jump in the write pointer causing a word to be skipped. Thus the control logic maintains the FIFO occupancy at the programmed delay value. ($C$). Figure \ref{coarse} shows the block diagram.
\subsubsection{Fine Delay}
\begin{figure}
\psfrag{Z}{$z^{-1}$}
\begin{center}
\pdfrackincludegraphics[width=10cm]{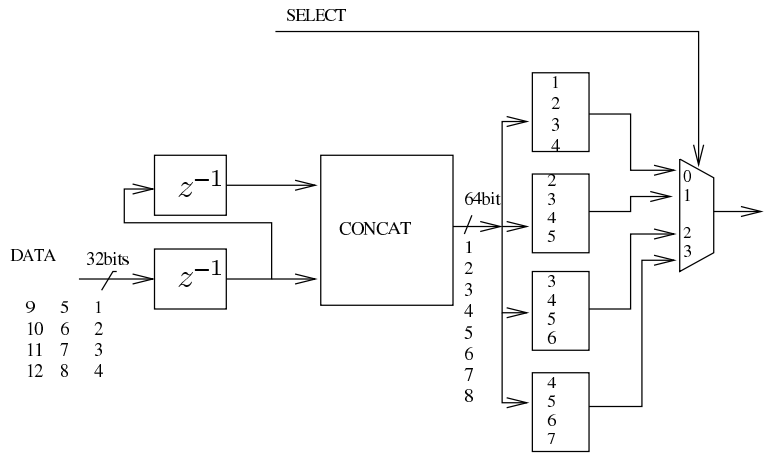}
\end{center}
\caption{\label{fine} Fine Delay is implemented using a barrel selector arrangement to re-align bytes.}
\end{figure}
The fine delay achieves a delay step size of $1$~ns. This stage simply performs a re-alignment of the quadruplet of samples arriving at one clock. This is done using a barrel selector arrangement shown in Figure \ref{fine}. It can be seen how we can select to realign the samples to sample number $2-3-4-5$ from the input of sample sequence $1-2-3-4$ by setting the select line to $1$. This shows that the select line value represents a phase advance of one sample rather than a delay. The fine delay block can adjust delays between $0$ and $3$ samples by setting the select line $S$ to $3-d$ where $d$ is the desired fine delay.  
\subsubsection{Super Fine Delay}
\begin{figure}
\begin{center}
\includegraphics[width=10cm]{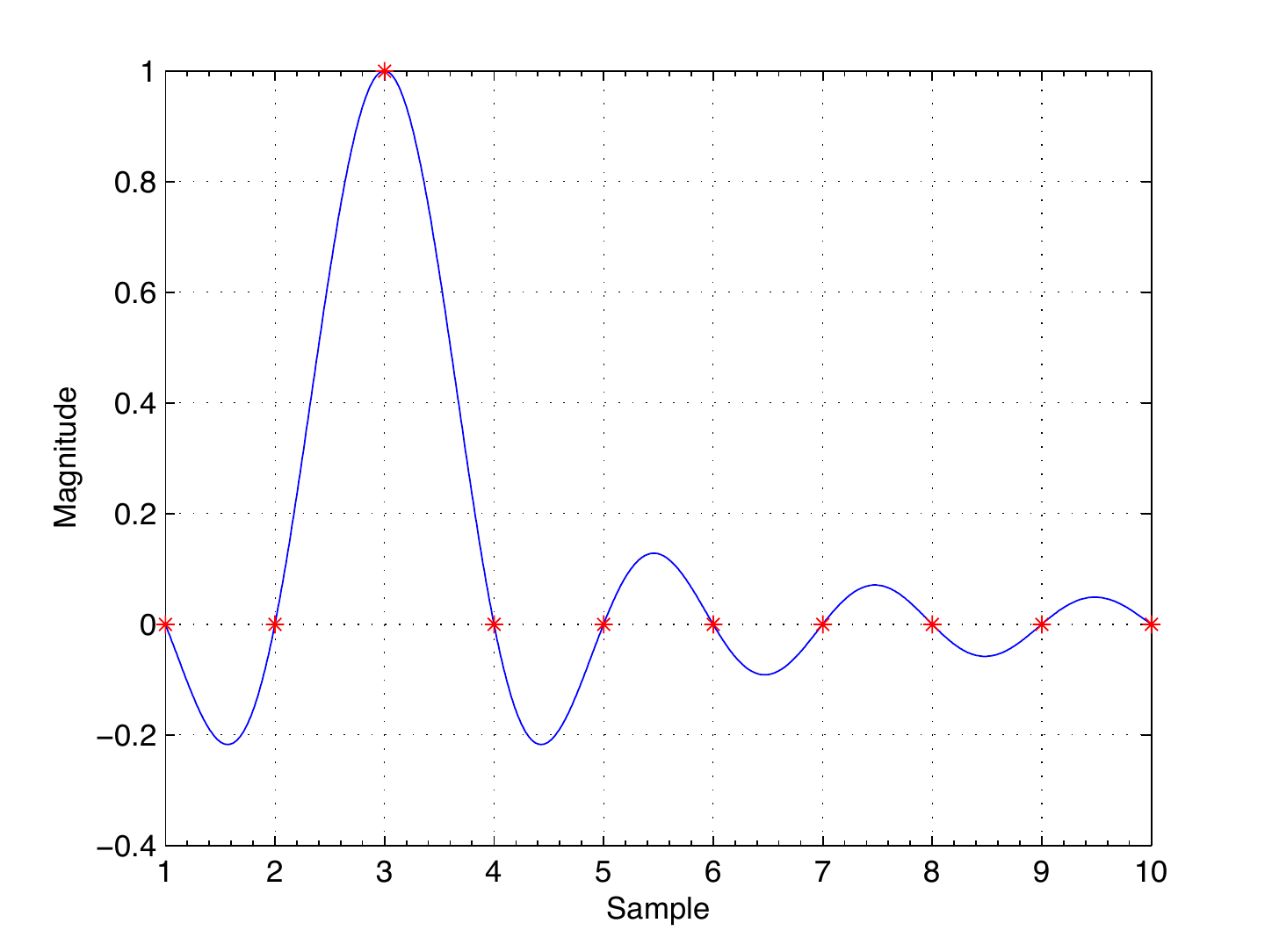}
\end{center}
\caption{\label{d3} Digital $10$ tap filter for $D=3$}
\end{figure}
\begin{figure}
\begin{center}
\includegraphics[width=10cm]{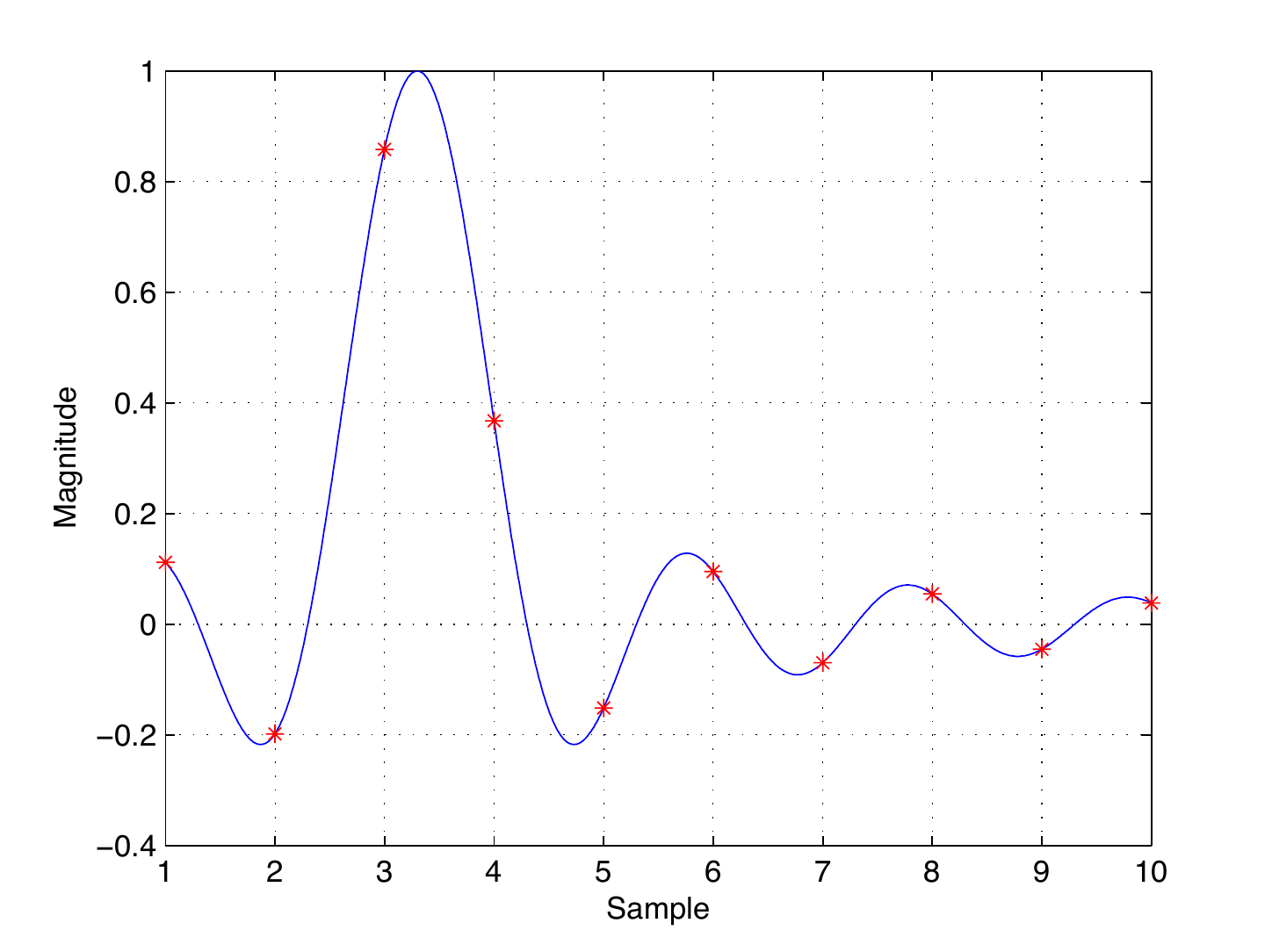}
\end{center}
\caption{\label{d33} Digital $10$ tap filter for $D=3.3$ showing asymmetrical FIR coefficients.}
\end{figure}
In the fine delay stage we could achieve the delay step equal to the sampling interval i.e. $\approx 1$~ns. To achieve a delay step finer than that would require to split the sampling interval and hence interpolate between samples. A delay line can be simply analyzed like a filtering operation. If the total desired delay is $D$ the digital filter output $y(n)$ can be written as 
\begin{equation}
y(n)=x(n-D)
\end{equation}
where $x(n)$ is the input sample stream. \cite{prinfd} In the $z$ transform domain the transfer function of this filter can be written as
\begin{equation}
H(z)=z^{-D}
\end{equation}
and the impulse response of this filter $h_D(n)$ can be shown to be,
\begin{equation}
h_D(n)=\frac{\sin \pi(n-D)}{\pi(n-D)}
\end{equation}

\begin{figure}
\psfrag{A}{$C_1$}
\psfrag{B}{$C_2$}
\psfrag{C}{$C_3$}
\psfrag{D}{$C_4$}
\psfrag{E}{$C_5$}
\psfrag{a}{$s_1$}
\psfrag{b}{$s_2$}
\psfrag{c}{$s_3$}
\psfrag{d}{$s_4$}
\psfrag{e}{$s_5$}
\psfrag{f}{$s_6$}
\psfrag{g}{$s_7$}
\psfrag{h}{$s_8$}
\psfrag{i}{$s_9$}
\psfrag{j}{$s_{10}$}
\psfrag{k}{$s_{11}$}
\psfrag{l}{$s_{12}$}
\psfrag{z}{$z^{-1}$}
\begin{center}
\pdfrackincludegraphics[width=10cm]{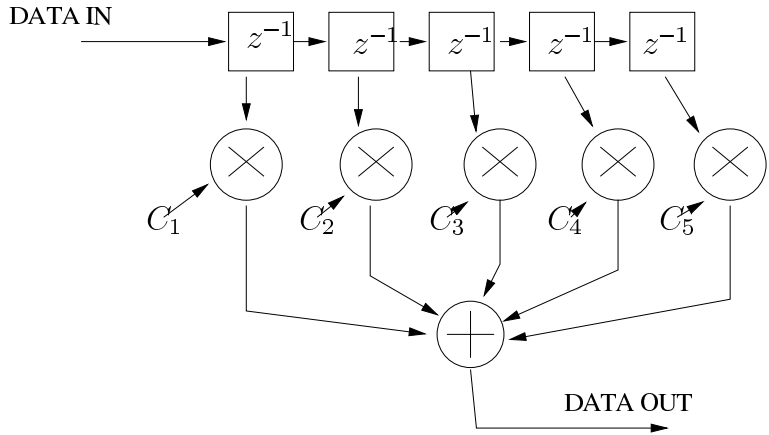}
\end{center}
\caption{\label{simple}Simple Finite Impulse Response Digital Filter Implementation}
\end{figure}
If $D$ is an integer number of sample periods, the impulse response of a $10$ tap filter simply corresponds to $3$ delay flip flops as can be seen in Figure \ref{d3} where $D=3$~samples. If $D$ is not an integer number of samples, say $D=3.3$~samples then the impulse response will be a shifted sinc pulse that is resampled as shown in Figure \ref{d33}. It is clear from these diagrams that a fractional delay filter cannot have symmetrical coefficients unless $D=0.5$. To implement a variable fractional delay line we have pre-computed coefficients for $10$ such filters corresponding to delays of $D=0.1$ to $D=0.9$ in steps of $0.1$~sample period and stored these coefficients in a RAM. We have implemented a $10$ tap digital FIR (Finite impulse response) filter in real time hardware and we can load any of these coefficient sets on demand using control logic. \\
Fig. \ref{simple} shows the digital implementation of a simple $5$ tap using a tapped delay line FIR filter where $C_1 ... C_5$ are the filter coefficients. However our system is demux-by-4 and we get 4 consecutive samples per clock. This requires a complex demux-by-4 pipelined FIR filter which we have designed.

\begin{figure}
\psfrag{c}{$C_n$}
\psfrag{s1}{$s^1_n$}
\psfrag{s2}{$s^2_n$}
\psfrag{s3}{$s^3_n$}
\psfrag{s4}{$s^4_n$}
\psfrag{s11}{$s^1_{n+1}$}
\psfrag{s12}{$s^2_{n+1}$}
\psfrag{s13}{$s^3_{n+1}$}
\psfrag{s14}{$s^4_{n+1}$}
\psfrag{x}{$\times$}
\psfrag{a}{$+$}
\psfrag{R}{Reorder}
\psfrag{P}{Pipeline Adjust}
\psfrag{p1}{$p^1_n$}
\psfrag{p2}{$p^2_n$}
\psfrag{p3}{$p^3_n$}
\psfrag{p4}{$p^4_n$}
\psfrag{p11}{$p^1_{n+1}$}
\psfrag{p12}{$p^2_{n+1}$}
\psfrag{p13}{$p^3_{n+1}$}
\psfrag{p14}{$p^4_{n+1}$}
\begin{center}
\pdfrackincludegraphics[width=10cm]{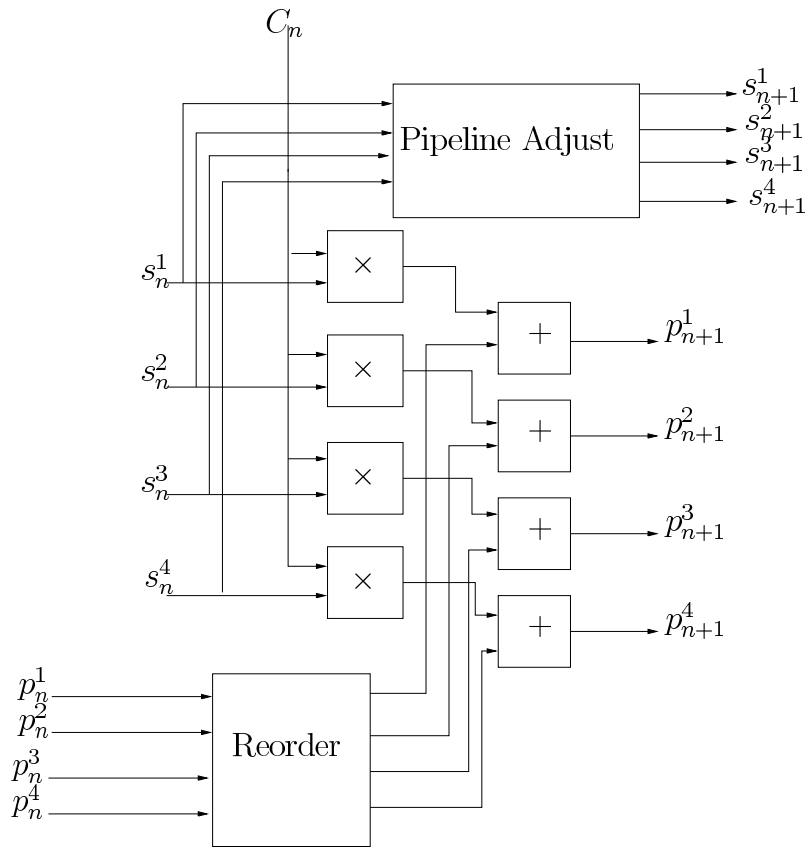}
\caption{\label{tap}Demux-by-4 FIR filter tap implementation}
\end{center}
\end{figure}

\begin{figure}
\psfrag{A}{$C_1$}
\psfrag{B}{$C_2$}
\psfrag{C}{$C_3$}
\psfrag{D}{$C_4$}
\psfrag{E}{$C_5$}
\psfrag{a}{$s_1$}
\psfrag{b}{$s_2$}
\psfrag{c}{$s_3$}
\psfrag{d}{$s_4$}
\psfrag{e}{$s_5$}
\psfrag{f}{$s_6$}
\psfrag{g}{$s_7$}
\psfrag{h}{$s_8$}
\psfrag{i}{$s_9$}
\psfrag{j}{$s_{10}$}
\psfrag{k}{$s_{11}$}
\psfrag{l}{$s_{12}$}
\psfrag{z}{$z^{-1}$}
\begin{center}
\pdfrackincludegraphics[width=10cm]{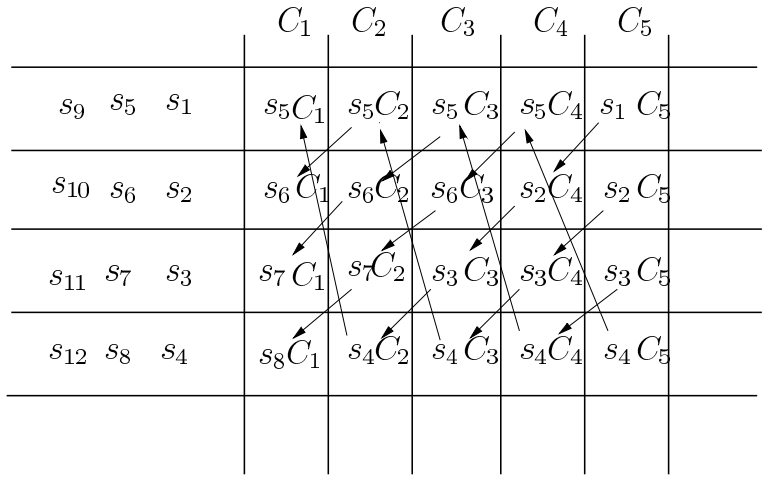}
\caption{\label{hairy}Demux-by-4 Finite Impulse Response Digital Filter Implementation}
\end{center}
\end{figure}
Such a filter is implemented by performing $4$ multiplications and $4$ partial sums per stage. Each stage corresponds to one filter tap. Partial sums are computed such that at any stage the output partial sums provide the correct FIR filtered output. Fig. \ref{tap} shows one tap of the filter chain. $p^1_n ... p^4_n$ are the partial sums and $s^1_n ... s^4_n$ the data samples passed on from the previous tap. The partial sums are reordered in everystage and the samples are delayed for adjusting the pipeline before the multiplication and addition operations are performed. Fig \ref{hairy} shows the $5$ tap case and how the first filtered output $o_1$ is computed in the pipeline.
\begin{equation}
o_1=s_1 \times C_5 + s_2 \times C_4 + s_3 \times C_3 + s_4 \times C_2 + s_5 \times C_1
\end{equation} 
\begin{figure}
\begin{center}
\begin{tiny}
\psfrag{D1}{$C_1$}
\psfrag{D2}{$C_2$}
\psfrag{D3}{$C_3$}
\psfrag{c1}{$C_1$}
\psfrag{c2}{$C_2$}
\psfrag{c3}{$C_3$}
\psfrag{F}{FIR Filter Pipeline}
\psfrag{s1}{$s_1$}
\psfrag{s2}{$s_2$}
\psfrag{s3}{$s_3$}
\psfrag{s4}{$s_4$}
\psfrag{U}{Update}
\psfrag{C}{Control}
\psfrag{z}{$z^{-1}$}
\psfrag{R}{RAM}
\pdfrackincludegraphics[width=14cm]{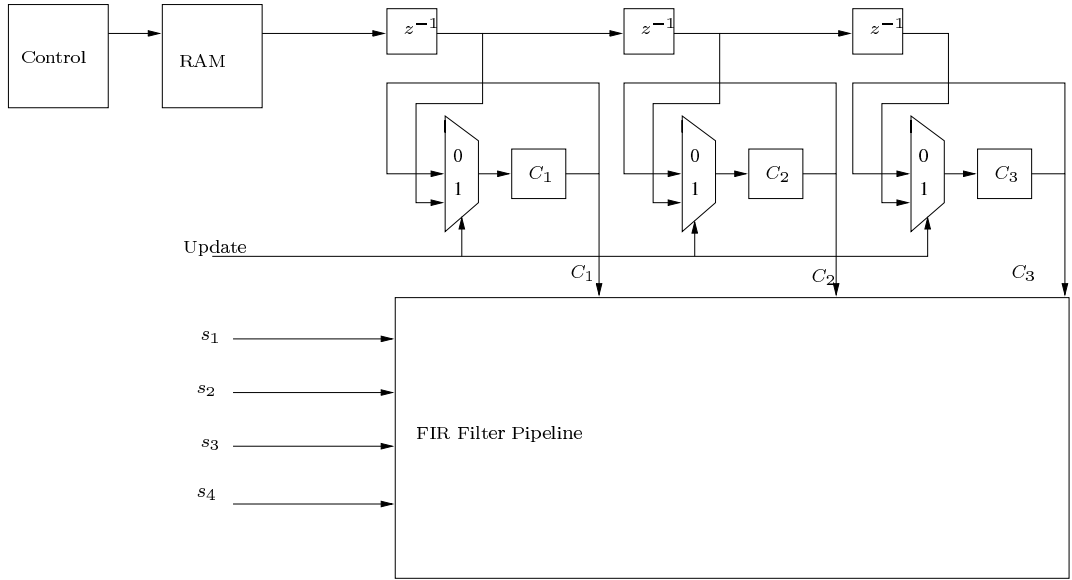}
\end{tiny}
\caption{\label{dbuffer}Double Buffering Scheme shown for a $3$ tap filter.}
\end{center}
\end{figure}
The filter coefficients can be changed and loaded from RAM to correspond to the desired fractional delay. A double buffering scheme is used such that the change of fractional delay can be put into effect instantaneously. The new set of coefficients is read out of RAM into a shift register which is used to parallel load the actual buffer which supplies the coefficients to the FIR pipeline. Fig. \ref{dbuffer} shows this arrangement for a $3$ tap case. When the delay needs to be changed the control logic reads out the corresponding set of new coefficients and when they have been read out into a serial load shift register an update pulse is sent out which loads the new set into the FIR pipeline.

\subsection{Delay Line Results}
\begin{figure}
\begin{center}
\includegraphics[width=14cm]{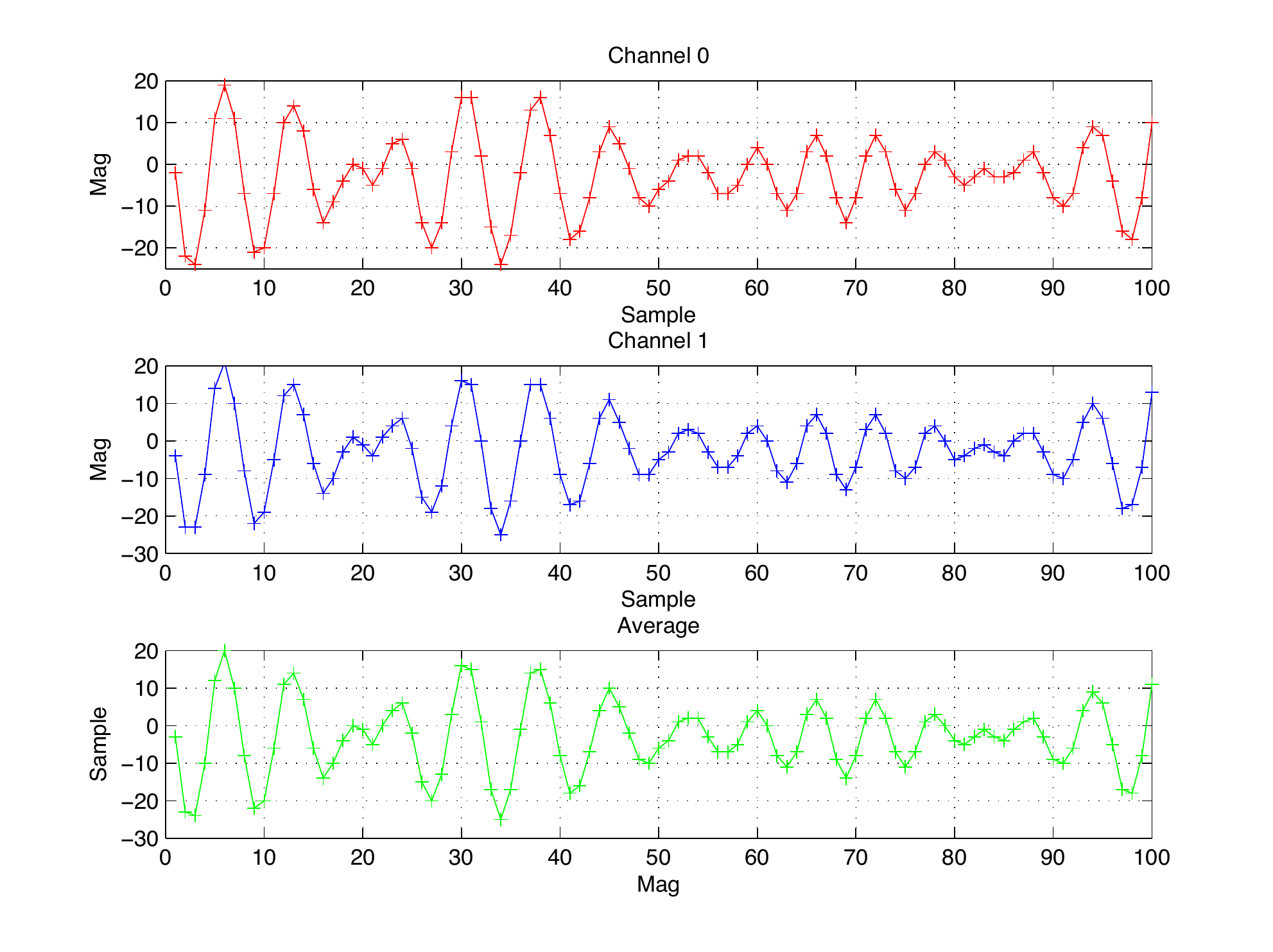}
\end{center}
\caption{\label{alligned}Snapshot of two data channels and their average when they are phase aligned}
\end{figure}

\begin{figure}
\begin{center}
\includegraphics[width=14cm]{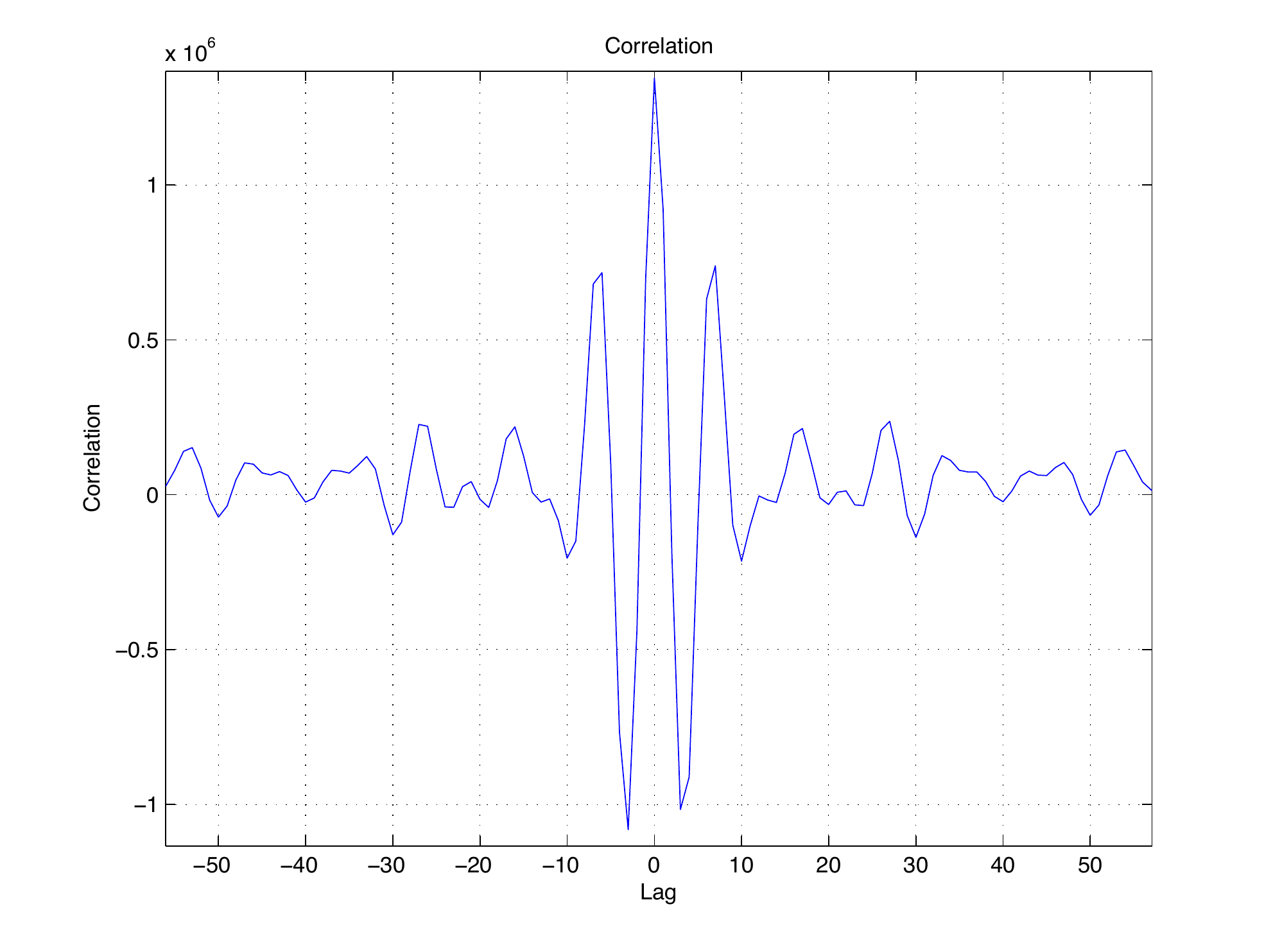}
\end{center}
\caption{\label{allignedcorr}Output of MATLAB function xcorr with $8000$ samples of $2$ aligned channels}
\end{figure}

The digital delay line can be tested by programming arbitrary delays into data streams and monitoring the output. Using the Power PC microprocessor we can record snapshots of data into onchip block RAMs. For this purpose we have designed a special circuit called \emph{iBOBscope}. The design simply uses a block RAM that is mapped to the Power PC bus with control logic to drive it address bus and write enable. We can store $\approx 8000$ time samples i.e. $8$~ns of data per channel using this scheme. The stored samples can be read into an external PC using RS-232 serial interface and then analyzed using MATLAB or similar tools. In our test setup we fed all the ADC inputs with the same band limited noise. The band limiting filter used was $100$~MHz wide thus highly oversampling the noise ($1024$~MHz) to see coherence in the various channels visually. Fig. \ref{alligned} shows the snapshot of $2$ data channels and their sum (actually average) when both ADC inputs were fed with the same signal through equal lengths of cable. It can be seen that the channels are phase aligned and their average is having the same total power as each of the channels. When the snapshots of these two channels are correlated with each other using MATLAB function \texttt{xcorr} the result shows a peak at $0$ indicating perfect phase alignment shown in Fig. \ref{allignedcorr}. We can now program our digital delay lines and then observe post-delay-line snapshots with the same test setup to check whether the delay lines are working correctly.

\begin{figure}
\begin{center}
\includegraphics[width=14cm]{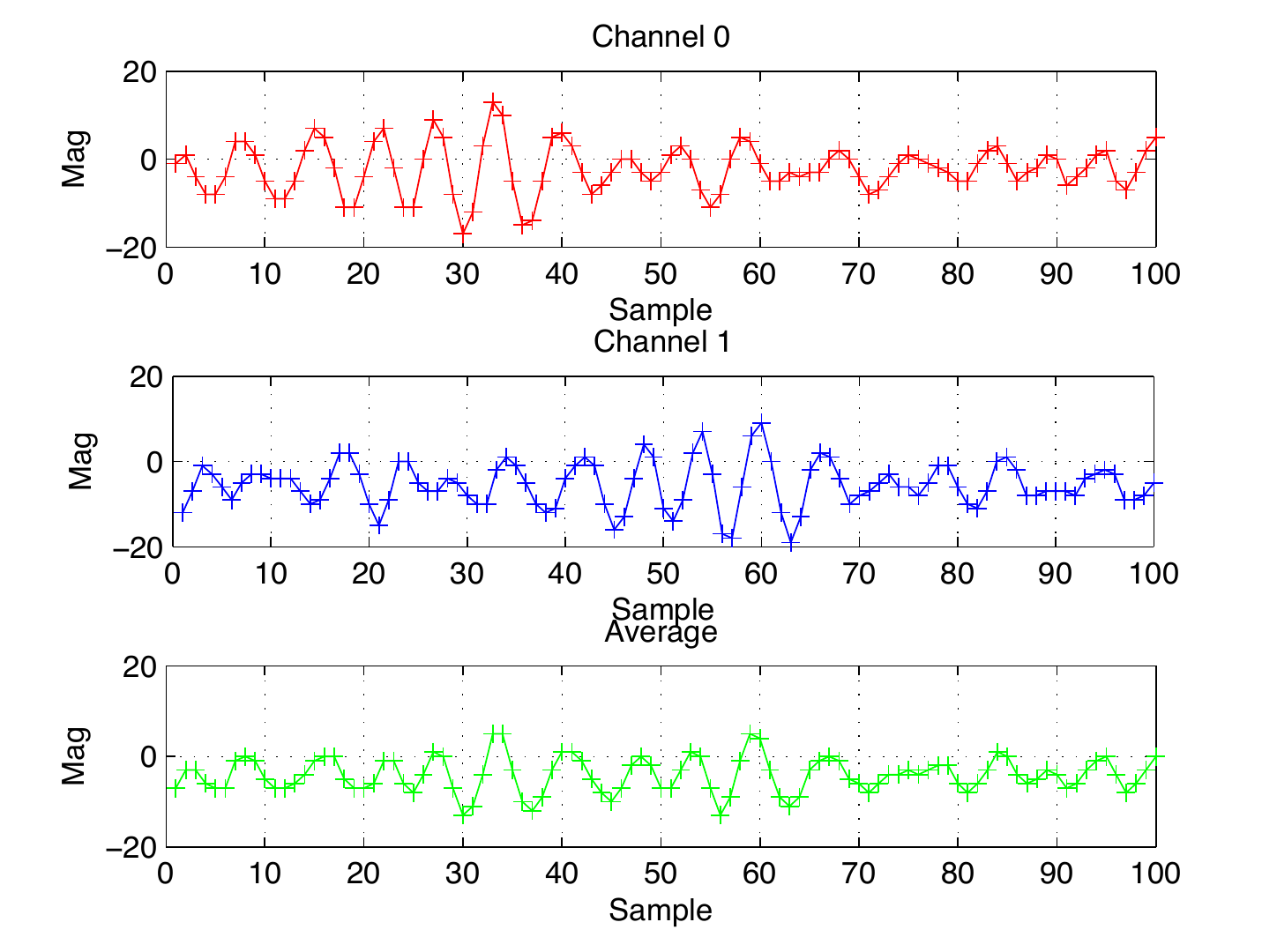}
\end{center}
\caption{\label{unaligned}Snapshot of two data channels and their average when one is delayed digitally by $26.5$~ns}
\end{figure}
\begin{figure}
\begin{center}
\includegraphics[width=14cm]{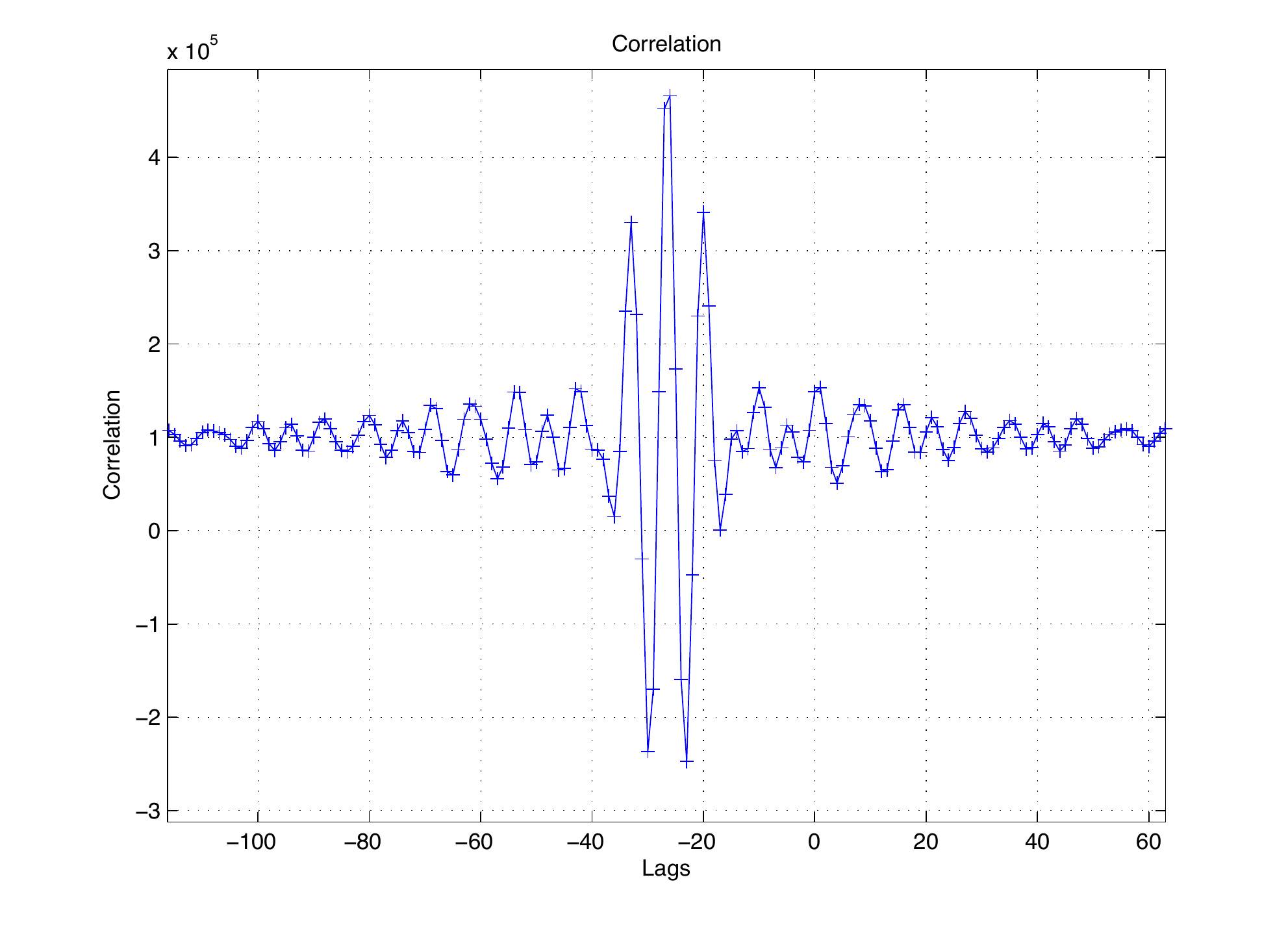}
\end{center}
\caption{\label{unalignedcorr}Output of MATLAB function xcorr with $8000$ samples of $2$ channels when one is delayed by $26.5$~ns}
\end{figure}
\begin{figure}
\begin{center}
\includegraphics[width=14cm]{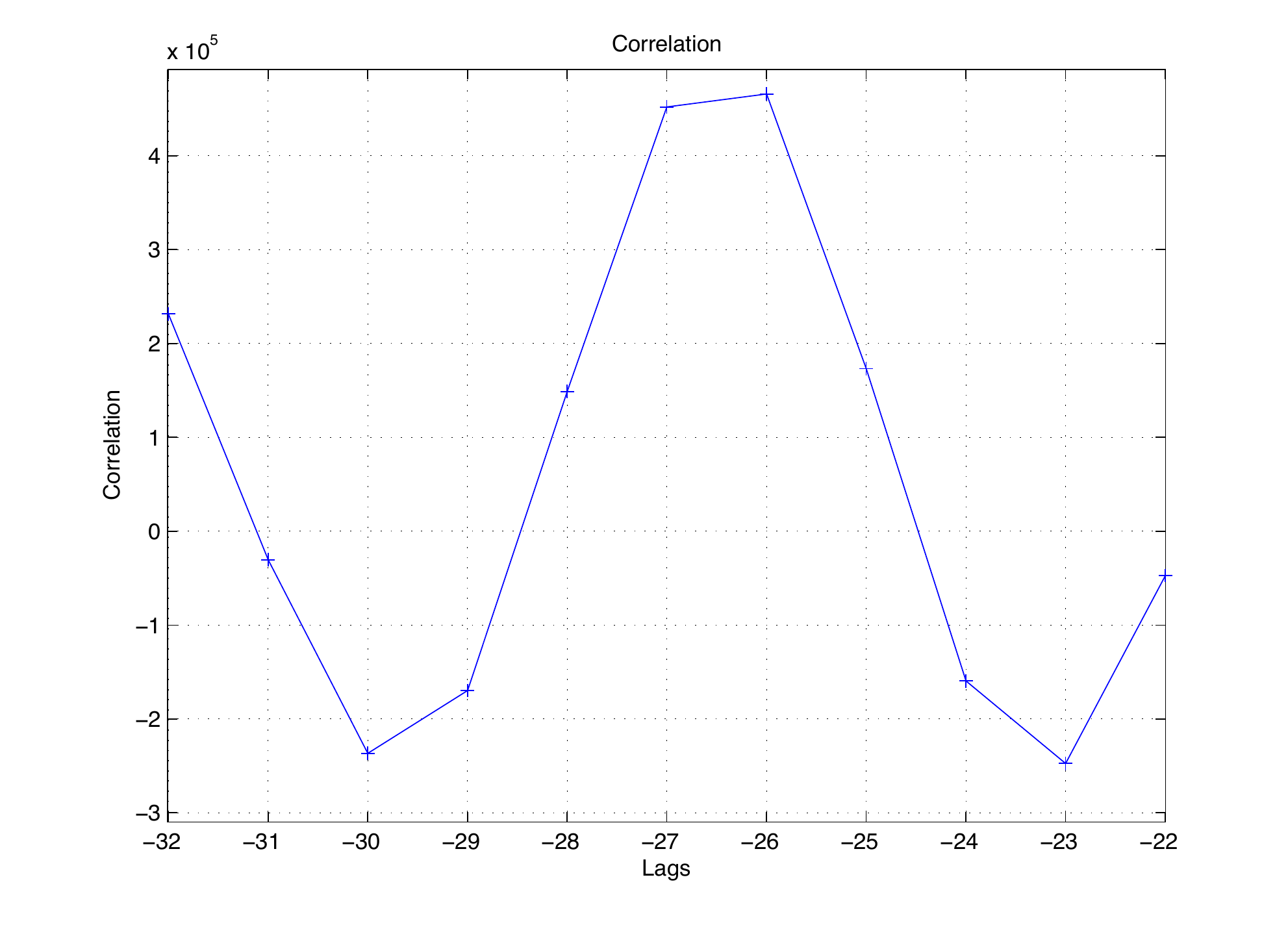}
\end{center}
\caption{\label{unalignedcorr1}Zoom into peak of Fig.\ref{unalignedcorr}}
\end{figure}

Fig. \ref{unaligned} shows the 2 channels and their average when one of the delay lines are programmed to a value of $26.5$~ns. We can observe that the channels are not phase aligned and the average signal power is reduced. We can repeat the MATLAB \texttt{xcorr} function on this set and see the correlation peak. As can be seen in Fig. \ref{unalignedcorr} and Fig. \ref{unalignedcorr1} the correlation peak is between $55$ and $56$~ns which assures us that the digital delay lines have correctly interpolated the samples for a delay of $26.5$~ns.

\section{Beamformer Calibration} 
Calibration implies the problem of finding the correct delay values to program into delay lines such that we get accurate phasing. 
The system delay can be classified into \emph{geometric, instrumental} and \emph{atmospheric} delays. The geometric delays can be calculated \emph{a priori} and the instrumental effects can be calibrated out by careful measurements along with some form of phase switching ($180^\circ$) arrangement. The atmospheric delays however change with time and need to be calibrated in real time for achieving good phase coherence. One way to calibrate out the atmospheric effects is to steer towards a strong astronomical source and use that as reference to find the atmospheric delay per antenna. However the astronomical signal is buried deep under system noise and we would need a correlator to measure the delays between channels. The SMA correlator can be run in parallel with the beamformer to extract this information.

\begin{figure}
\psfrag{a}{$\tau_1$}
\psfrag{b}{$\tau_2$}
\psfrag{c}{$\tau_3$}
\psfrag{d}{$\tau_4$}
\psfrag{e}{$\tau_5$}
\psfrag{f}{$\tau_6$}
\psfrag{g}{$\tau_7$}
\psfrag{h}{$\tau_8$}
\begin{center}
\pdfrackincludegraphics[width=12cm]{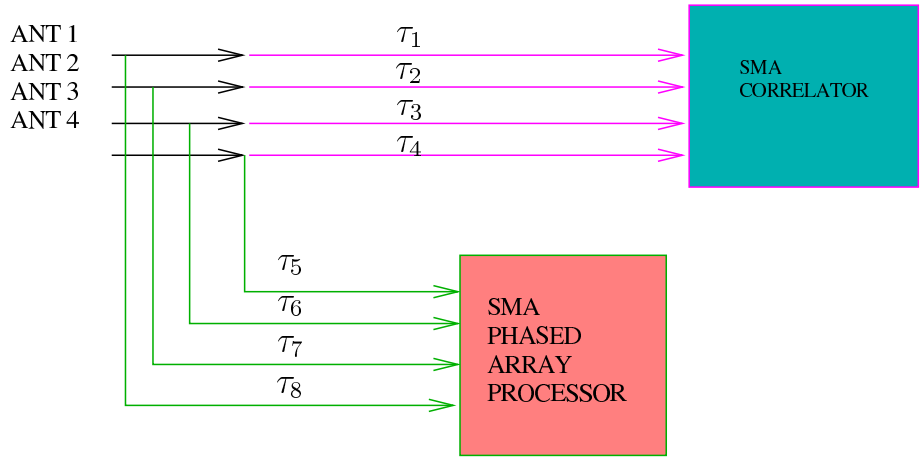}
\end{center}
\caption{\label{corr} Signal flow to SMA correlator and phased array processor}
\end{figure}
The signal path to the SMA correlator follows a different analog path (and analog processing chain) than the SMA beamformer as seen in fig. \ref{corr}. The differences in analog paths in various antennas $\tau_1, \tau_2 ..$ etc. must be calibrated before the SMA correlator can be used to continuously track the atmosphere. To solve this problem it was decided to reuse a correlator being built by CASPER on the iBOB board. It was initially decided to just use this correlator for calibrating the SMA but it eventually turned out that we had to redesign some parts of the CASPER correlator to suit the requirements of our project. It was then decided that we can use this custom designed \emph{Calibration Correlator} (using CASPER library components) for calibrating the SMA beamformer independent of the SMA correlator.
\begin{figure}
\begin{center}
\includegraphics[width=12cm]{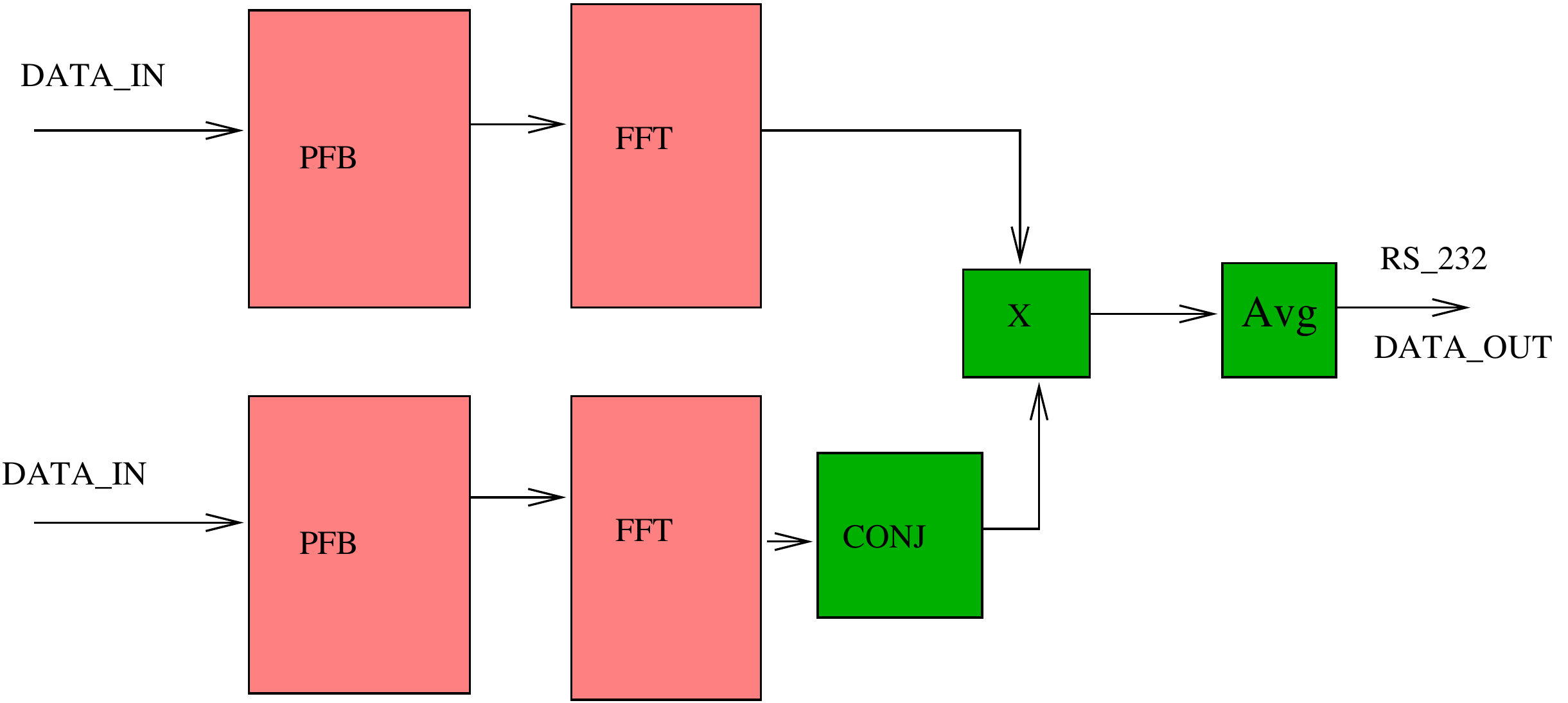}
\end{center}
\caption{\label{corr1} Architecture of \emph{Calibration Correlator}}
\end{figure}
\subsection{Design of Calibration Correlator} 
To accurately phase up 8 antennas we need to know 7 delays, i.e. delay of each antenna w.r.t. a reference antenna. If we can measure the cross-correlations of each of 7 antennas w.r.t the reference we can derive these delays from the phase slope in correlation functions. If we observe a fairly strong astronomical source and measure these 7 correlation functions for a few seconds of integration we would be able to track these calibration delays. Since atmospheric and instrumental effects (even geometric effects) do not change significantly over the order of a few minutes we do not need to measure these 7 correlations simultaneously. We should be able to use an easy-to-design single baseline correlator and time multiplex these 7 measurements on it.
 The CASPER astronomy library provides us with pre-designed blocks for FIR based implementation of poly-phase filter banks (PFB) and Fast Fourier transform (FFT) circuits. The PFB performs a pre-conditioning of signal before computing FFT to get a sharper frequency resolution. The PFB can be shown to have better performance than windowing and overlap methods commonly used with FFT. Using these blocks we added circuitry to multiply two FFT streams to get the cross correlation function and integrate the result for a few seconds. Fig. \ref{corr1} shows a system diagram of the designed correlator. It is an $FX$ correlator with $64$ complex/$32$ real frequency channels and can do onchip integration of about $16$~seconds. The correlation spectra can be read out using RS-232 serial interface to a control PC. The design heavily relies on the Polyphase Filter Bank (PFB) and Fast Fourier Transform (PFB-FFT) library developed by CASPER. The PFB-FFT provide a highly optimized implementation of the FFT algorithm with preceding decimating FIR filters to give sharp spectral response. \cite{aaron} After the PFB-FFT stage we have implemented a complex multiply (with complex conjugate) operation to give correlation result. A Vector Accumulator (Avg) has been designed to integrate the spectra for a maximum of $16$ seconds in Block RAMs. The integration result is stored in CPU mapped memory to read out via serial port. Fig. \ref{corr1} shows the block diagram of the correlator with green blocks representing ones designed by us and pink ones supplied by the CASPER library.
The design of this correlator consumed a very large portion of the project time line because of the challenges involved and the fact that the CASPER libraries were undergoing constant development and upgrades during our project. The various design challenges we faced included dynamic range, sensitivity and fitting the logic within the FPGA with the specified timing. ($256$~MHz clock rate) 
We have tested the operation of the correlator using a test setup wherein a correlated component (simulated $480$~MHz wideband noise) was added to two independent simulated receiver noise components ($480$~MHz uncorrelated noise sources).  
\subsection{Calibration Correlator Results}
\begin{figure}
\begin{center}
\includegraphics[width=14cm]{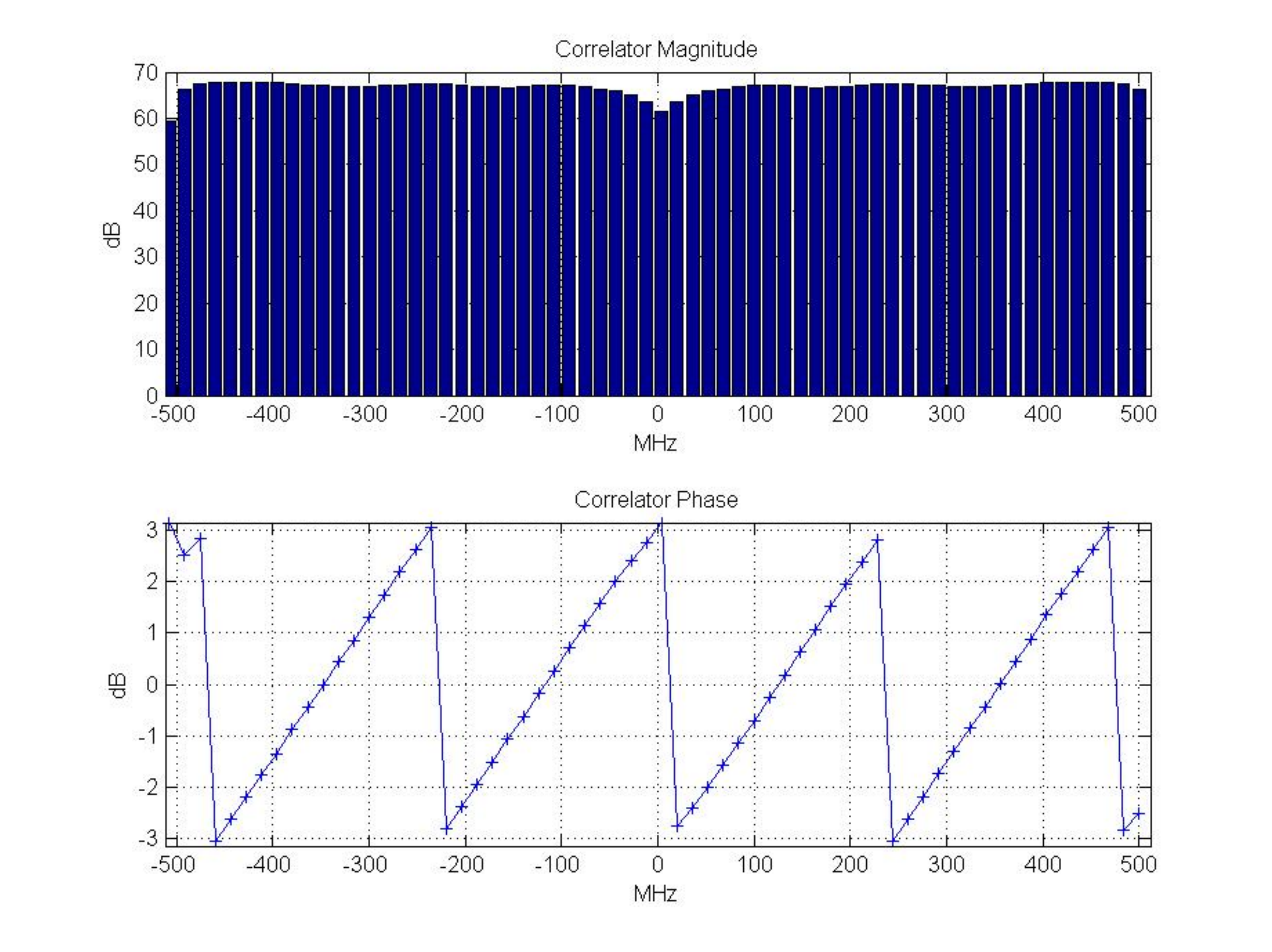}
\end{center}
\caption{\label{auto} Autocorrelation Spectra ($\infty$ Signal to Noise)}
\end{figure}

\begin{figure}
\begin{center}
\includegraphics[width=14cm]{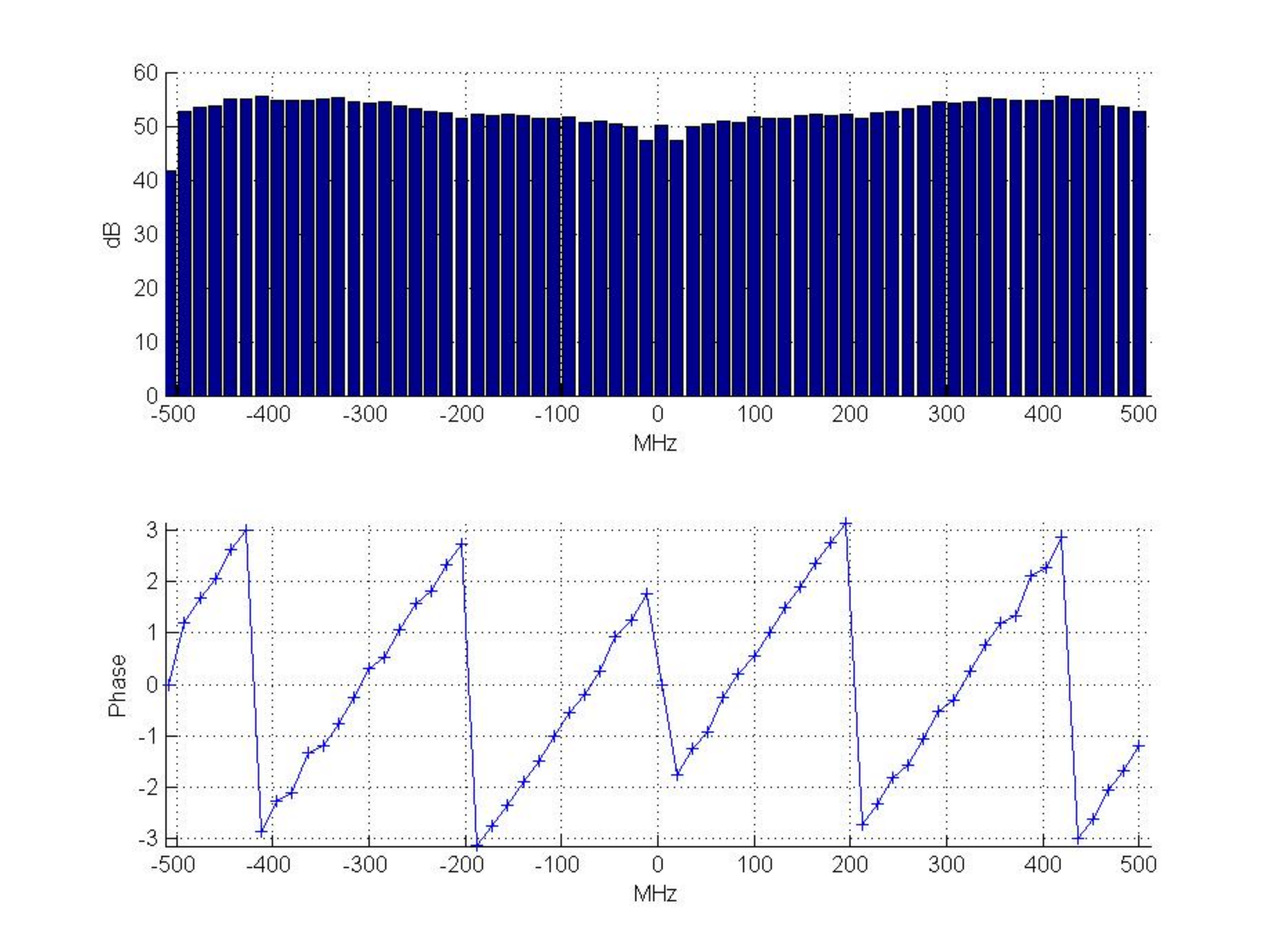}
\end{center}
\caption{\label{snr2} Cross correlation Spectra with $-2$~dB Signal to Noise}
\end{figure}

\begin{figure}
\begin{center}
\includegraphics[width=14cm]{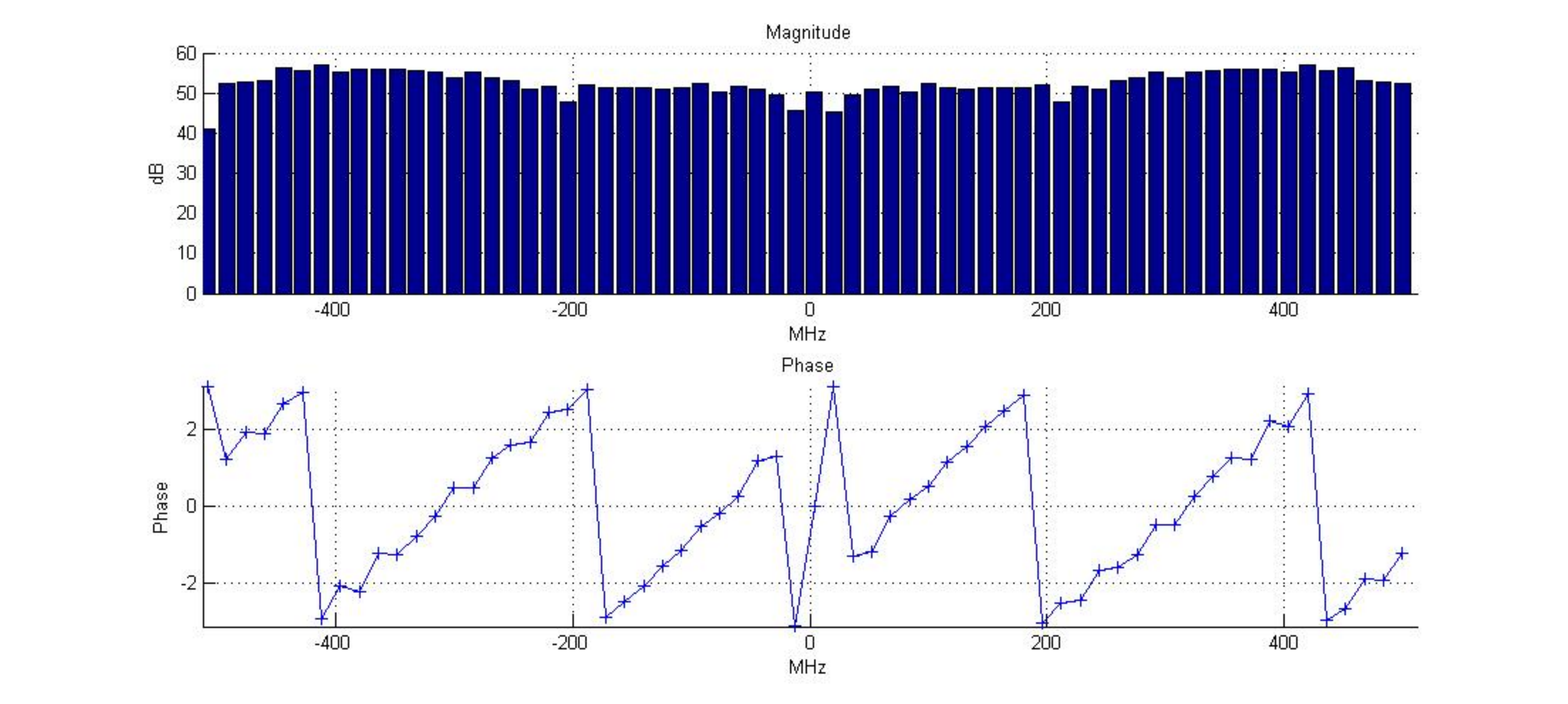}
\end{center}
\caption{\label{snr9} Cross correlation Spectra with $-9$~dB Signal to Noise}
\end{figure}

\begin{figure}
\begin{center}
\includegraphics[width=14cm]{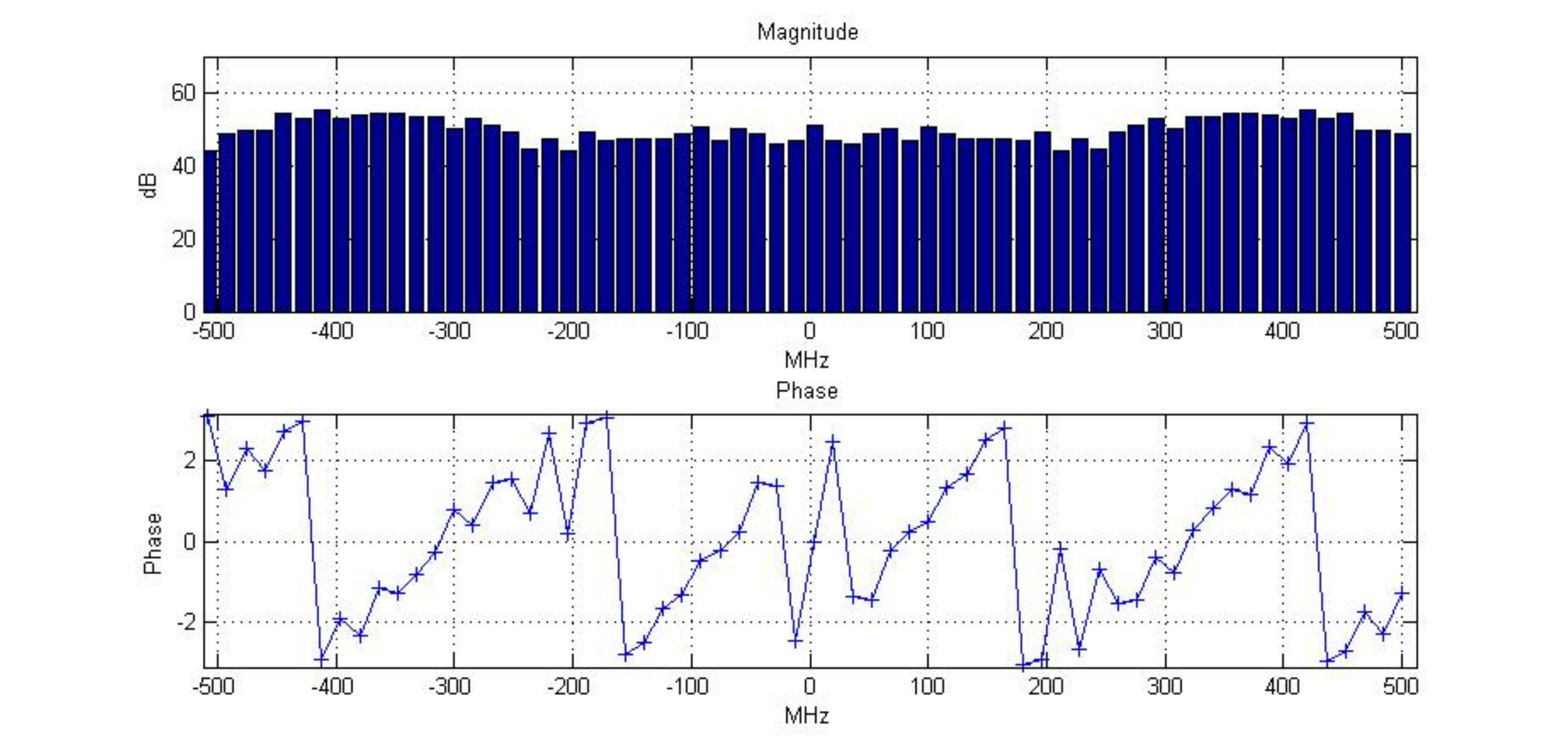}
\end{center}
\caption{\label{snr12} Cross correlation Spectra with $-9$~dB Signal to Noise}
\end{figure}

\begin{figure}
\begin{center}
\includegraphics[width=14cm]{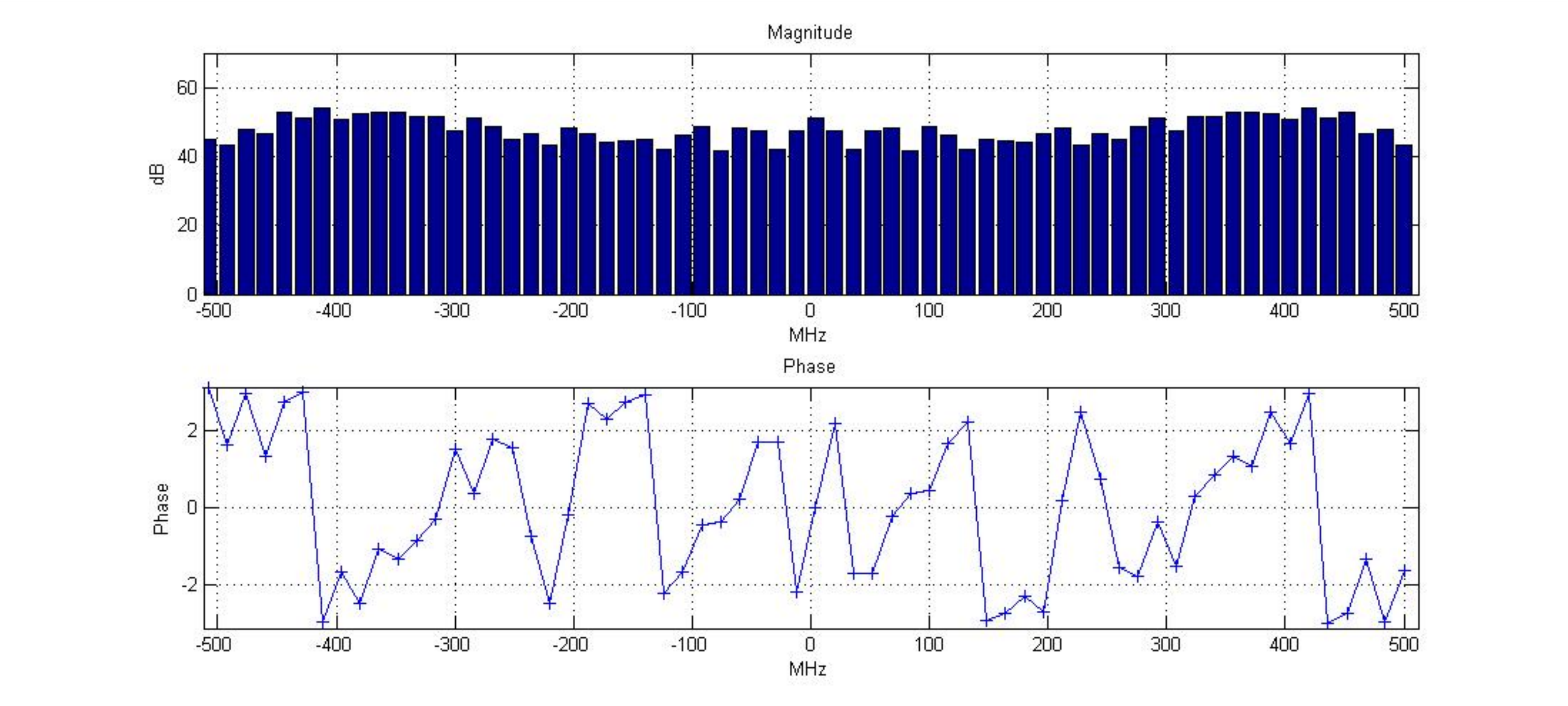}
\end{center}
\caption{\label{snr15} Cross correlation Spectra with $-9$~dB Signal to Noise}
\end{figure}

\begin{figure}
\begin{center}
\includegraphics[width=14cm]{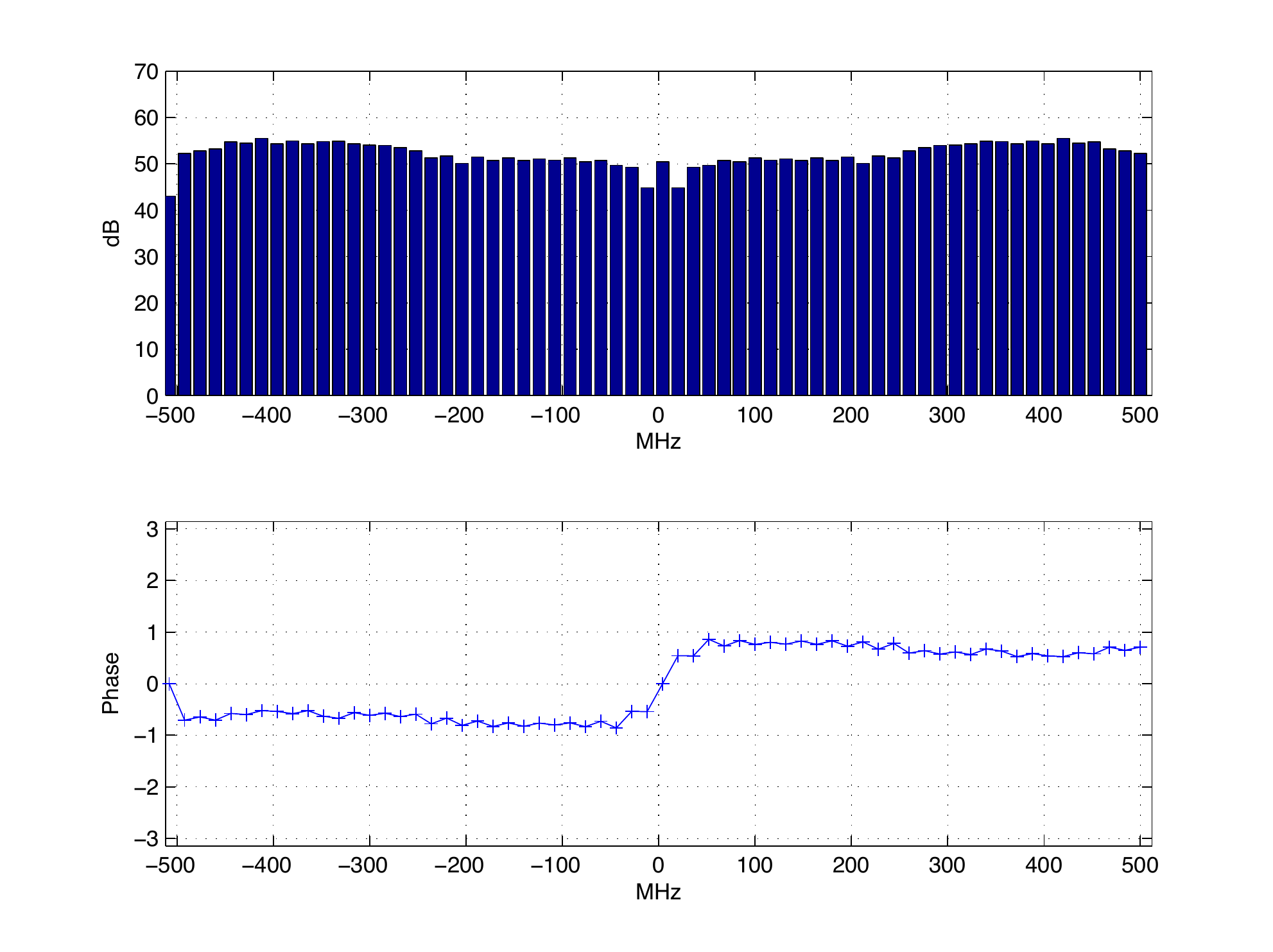}
\end{center}
\caption{\label{nodelay} Cross correlation Spectra with no cable delay}
\end{figure}

Fig. \ref{auto} shows auto correlation spectra measured with the designed correlator. As we can notice for real data we get a two sided spectrum. The phase plot shows a slope corresponding to the difference in cable lengths feeding the 2 ADC inputs. Fig. \ref{snr2}, \ref{snr9}, \ref{snr12}, \ref{snr15} show the cross correlation measured using progressively weaker correlated component signals. Fig. \ref{nodelay} shows the cross correlation when the cable lengths are equalized thereby giving a flat phase response.

\subsection{Final Architecture}
\begin{figure} 
\includegraphics[scale=0.65]{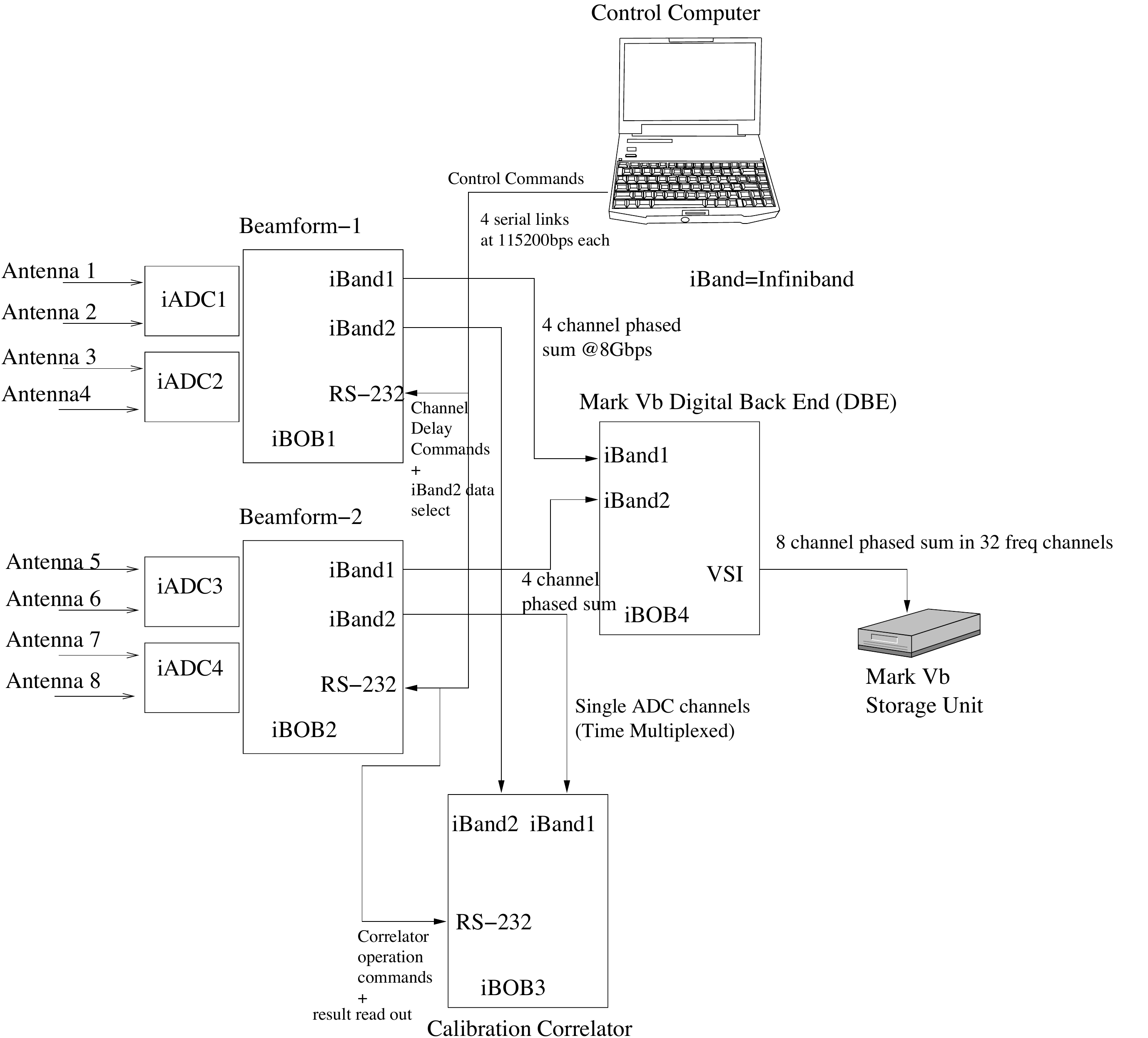}
\caption{\label{arch} Overall System Architecture}
\end{figure}

\begin{figure}
\psfrag{A}{M5 Recorder}
\psfrag{B}{DBE}
\psfrag{T}{VSI}
\psfrag{C}{VLBI In}
\psfrag{U}{iBOB-1}
\psfrag{V}{iBOB-2}
\psfrag{P}{Ant 1}
\psfrag{Q}{Ant 2}
\psfrag{R}{Ant 3}
\psfrag{S}{Ant 4}
\psfrag{E}{Ant 5}
\psfrag{F}{Ant 6}
\psfrag{G}{Ant 7}
\psfrag{H}{Ant 8}
\psfrag{X}{$8$~Gbps}
\psfrag{Y}{$8$~Gbps}
\psfrag{Z}{$\approx8$~Gbps per antenna}
\psfrag{G}{Correlator}
\begin{center}
\pdfrackincludegraphics[width=12cm]{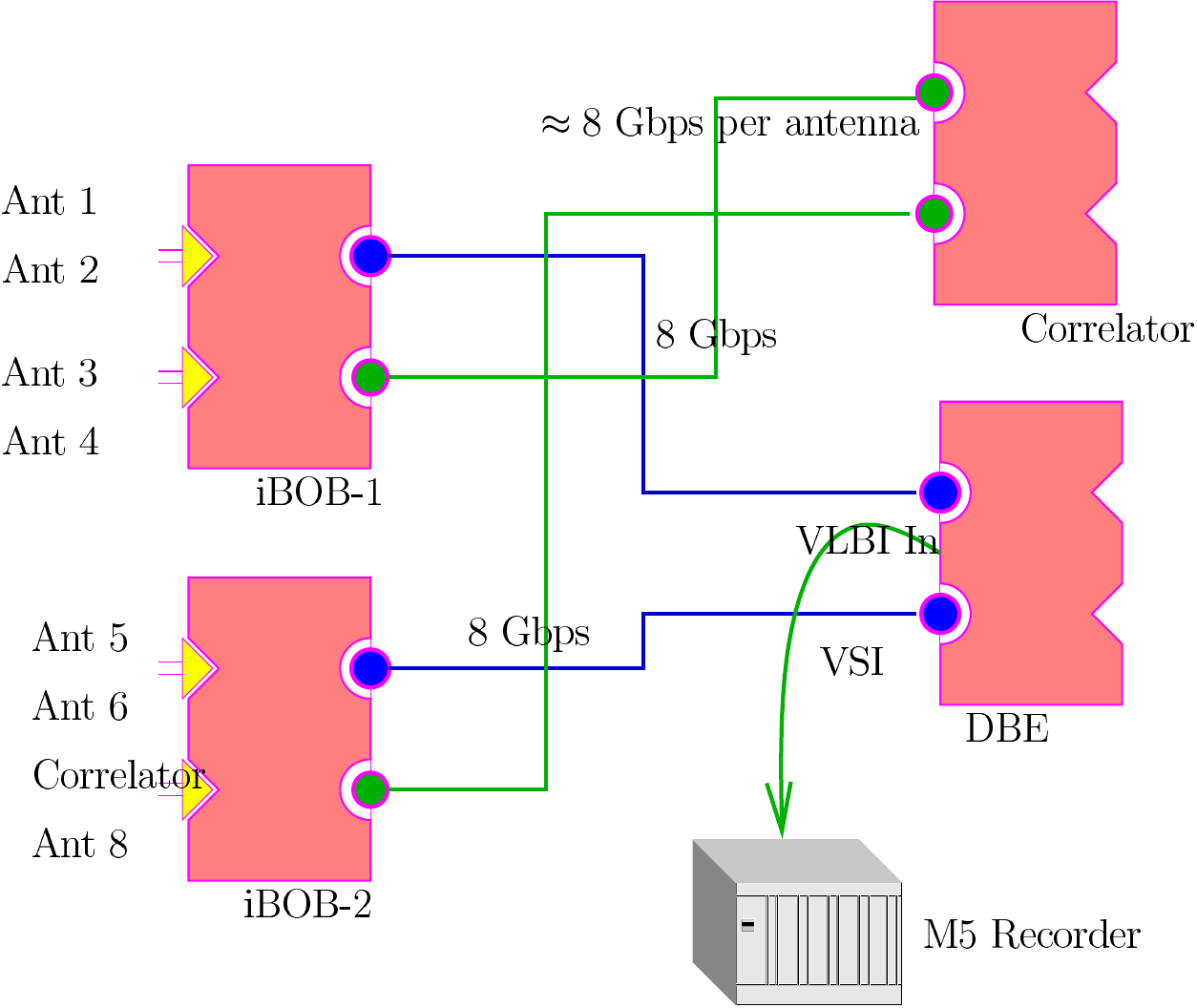}
\end{center}
\caption{\label{final} Final Architecture}
\end{figure}

Finally the correlator with time multiplexed correlation calibrations can be seen in Fig. \ref{arch} and Fig. \ref{final}. 
 The spare Infiniband connector on \emph{Beamform} boards described in (1) is used to spool the sampled and delayed data channels (before sum) through the spare Infiniband in a time multiplexed fashion. The \emph{Correlator} receives 2 channels from the \emph{Beamform} boards over XAUI and calculates correlation spectra which are read out over RS-232 to the control computer which can extract delays from it and appropriately program the delay lines in \emph{Beamform} boards.

\section{Discussion and Future Work}
The system built during this masters project has cleared away most of the difficult problems in the path of a working beam former on the Mauna Kea summit. However some components need some more work before the system can be ready for a sky test. 
\begin{enumerate}
\item \textbf{Correlator Sensitivity} As seen in the previous sections the calibrating correlator cannot extract delays with reasonable confidence if the correlated signal goes more than $15$~dB below noise. With an astronomical signal the expected SNR is about $-30$ to $-40$~dB. Some work is needed to improve the sensitivity of the correlator.
\item \textbf{XAUI Links} Though we have tested XAUI links and their synchronization, we have not integrated these links with the above designs. It might be needed to add additional circuitry to designs to take care of infrequent link failures in the XAUI.
\item \textbf{Mark $5b$ DBE} Integration blocks for the Mark V DBE have been designed but have not yet been tested with actual Mark $5b$ recorders because of logistic problems with acquiring them.
\item \textbf{Delay Extraction and Automation} Software modules to extract delays automatically from the correlator result and program delays lines need to be built.
\item \textbf{Fringe Rotation and Analog Subsystem} We need to have a primary and secondary phase rotating sample clock to correct for fringe rotation. This analog subsystem is to be built and tested. 
\end{enumerate}

In conclusion we have demonstrated the power of FPGA based back end design in this masters project. With a very short development cycle FPGA's with supporting infrastructure can help us build telescope backends. We have also come a long way with the design of a phased array processor for the telescopes on the Mauna Kea summit. 

\bibliographystyle{plain}
\bibliography{outline.bib}
\end{document}